# Geo-neutrinos and the Radioactive Power of the Earth

S.T. Dye
College of Natural Sciences, Hawaii Pacific University, Kaneohe, HI, 96744 USA
Department of Physics and Astronomy, University of Hawaii, Honolulu, HI, 96822 USA

**Abstract**
Chemical and physical Earth models agree little as to the radioactive power of the planet. Each predicts a range of radioactive powers, overlapping slightly with the other at about 24 TW, and together spanning 14 – 46 TW. Approximately 20 % of this radioactive power (3 – 8 TW) escapes to space in the form of geo-neutrinos. The remaining 11 – 38 TW heats the planet with significant geo-dynamical consequences, appearing as the radiogenic component of the 43 – 49 TW surface heat flow. The non-radiogenic component of the surface heat flow (5 – 38 TW) is presumably primordial, a legacy of the formation and early evolution of the planet. A constraining measurement of radiogenic heating provides insights to the thermal history of the Earth and potentially discriminates chemical and physical Earth models. Radiogenic heating in the planet primarily springs from unstable nuclides of uranium, thorium, and potassium. The paths to their stable daughter nuclides include nuclear beta decays, producing geo-neutrinos. Large sub-surface detectors efficiently record the energy but not the direction of the infrequent interactions of the highest energy geo-neutrinos, originating only from uranium and thorium. The measured energy spectrum of the interactions estimates the relative amounts of these heat-producing elements, while the intensity estimates planetary radiogenic power. Recent geo-neutrino observations in Japan and Italy find consistent values of radiogenic heating. The combined result mildly excludes the lowest model values of radiogenic heating and, assuming whole mantle convection, identifies primordial heat loss. Future observations have the potential to measure radiogenic heating with better precision, further constraining geological models and the thermal evolution of the Earth. This review presents the science and status of geo-neutrino observations and the prospects for measuring the radioactive power of the planet.

**1. Introduction**
Heat flows from the surface of the Earth at a recently estimated rate of 47 ±2(stat) TW [*Davies and Davies*, 2010]. This result, which is based on more than 38,000 observations of heat conduction in surface rock and adjusted for hydrothermal circulation and volcanism, is consistent with previous estimates of 46 ±3 TW [*Jaupart et al.*, 2007] and 44 ±1 TW [*Pollack et al.*, 1993], using a subset of the same data. These concurring estimates establish a power budget for the Earth in the range of 43 – 49 TW. A relatively unconstrained portion of the surface heat flow is primordial, originating from the formation and early evolution of the planet. Several potential sources of primordial heat each liberate enough energy to exceed the present surface heat flow, when averaged over the 4.5 Gy lifetime of the Earth. These sources include gravitational energy released by the segregation of a metallic iron-nickel core from a homogeneous proto Earth [*Birch*, 1965; *Shaw*, 1978], nuclear energy from the relatively rapid decay of the original inventory of $^{26}$Al [*MacPherson et al.*, 2012], and nuclear energy from the much slower decays of $^{238}$U, $^{235}$U, $^{232}$Th, and $^{40}$K [*Van Schmus*, 1995]. The loss of this fossil heat, which is the portion of the surface heat flow not due to contemporary radioactivity, determines the rate of planetary cooling [*Verhoogen*, 1980]. Estimates of heat entering the mantle from the core vary from 5 – 15 TW [*Lay et al.*, 2008]. Estimates of mantle cooling, derived from continental uplift [*Galer and Mezger*, 1998], geochemistry of dike swarms [*Mayborn and Lesher*, 2004], and isotopic analysis of xenoliths [*Bedini et al.*, 2004], vary from 7 – 15 TW. Together these estimates propose the planet is losing fossil heat at a rate of 12 – 30 TW, suggesting 13 – 37 TW of contemporary radiogenic heating.

Radiogenic heating derives primarily from nuclides of uranium, thorium, and potassium [*Van Schmus*, 1995]. Typical terrestrial ratios of these elements (Th/U≈4 and K/U≈10$^4$) project dominant contributions to radiogenic heating (~40% each) from uranium and thorium, with potassium supplying the remainder. These elements condense as oxides, giving them chemical affinity for silicate minerals. Accordingly, all geological models discussed herein assume the metallic core is effectively non-radioactive [*McDonough*, 2003], limiting radiogenic heating to the primitive mantle (PM), which is synonymous with the bulk silicate Earth. Two general classes of geological models, cosmochemical (CC) and geophysical (GP), predict different ranges of radiogenic heating in the bulk silicate Earth. The ranges overlap slightly at about 20 TW and do not exceed the 43 – 49 TW surface heat flow. Cosmochemical models predict the low range of radiogenic heating, allowing for a higher rate of planetary cooling than geophysical models, which predict the high range of radiogenic heating.

Cosmochemical models predict U and Th abundances in primitive mantle enriched relative to the abundances of these refractory elements in chondritic meteorites. These models estimate the primitive mantle abundance of moderately volatile K relative to U, typically setting K/U (≈ 12,000) consistent with a recent finding of the terrestrial ratio [*Arevalo et al.*, 2009]. Assuming the primitive mantle has chondritic abundances of U and Th ($a_U$=7.8 ppb ±10 %; $a_{Th}$=29.8 ppb ±10 %) [*Palme and O'Neill*, 2003] and an abundance of K ($a_K$=94 ppm ±10 %) from the terrestrial K/U ratio predicts radiogenic heating of ≈7.5 TW, which is comparable to the estimated contribution from continental crust [*Rudnick and Gao*, 2003; *Jaupart and Mareschal*, 2004]. Radiogenic heating in the present-day mantle requires the primitive mantle to have abundances of refractory heat-producing elements enriched compared with chondrites. Accordingly, chondritic U and Th abundances establish a convenient reference for comparing model predictions of primitive mantle abundances. The minimum enrichment predicted by a cosmochemical model is ≈1.5 [*Javoy et al.*, 2010], corresponding to primitive mantle radiogenic heating of ≈11 TW. Enrichment of ≈1.5 is consistent with segregation of a non-radioactive Earth core from a homogeneous proto planet. Cosmochemical models typically specify U and Th in the bulk silicate Earth enriched relative to chondritic abundances by 2.5 to a maximum of 2.8 [*Hart and Zindler*, 1986; *McDonough and Sun*, 1995; *Palme and O'Neill*, 2003], significantly more than expected from core segregation alone. However, a new statistical analysis recommends intermediate values of 1.8 – 2.5 [*Lyubetskaya and Korenaga*, 2007]. The maximum enrichment corresponds to primitive mantle radiogenic heating of ≈21 TW, while the intermediate values correspond to 13 – 18 TW. Cosmochemical models predict a range of radiogenic heating in the bulk silicate Earth, spanning 11 – 21 TW.

Geophysical models use a scaling law to relate mantle convection to heat loss, predicting the thermal evolution of the Earth. Refining the description of convection, mainly focusing on mantle viscosity, and tuning the strength of this relationship selects solutions consistent with estimates of previous and present conditions. Mantle viscosity depends exponentially on temperature, which is influenced by radiogenic heating. The ratio of mantle radiogenic heating to mantle heat flow defines the convective Urey ratio [*Korenaga,* 2008]. Setting the heat flow out of the mantle to ≈36 TW (surface heat flow minus crustal heat production) normalizes the comparison of geophysical models as characterized by the convective Urey ratio. The traditional parameterized convection model proposes a convective Urey ratio ≈0.8 [*Turcotte,* 1980]. This ratio establishes the maximum predicted by a geophysical model, corresponding to primitive mantle radiogenic heating of ≈38 TW. Accounting for variable viscosity convection finds a somewhat lower convective Urey ratio of ≈0.5 [*Christensen,* 1985], corresponding to primitive mantle radiogenic heating of ≈27 TW. Including the effects of water on mantle viscosity in addition to temperature predicts a convective Urey ratio ≈0.3 [*Crowley et al.*, 2011]. This ratio, establishing the minimum predicted by a geophysical model, corresponds to primitive mantle radiogenic heating of ≈19 TW. Geophysical models predict a range of radiogenic heating in the bulk silicate Earth, spanning 19 – 38 TW.

The heat flow out of the mantle is approximately the difference of the surface heat flow and the radiogenic heating in the crust. Seismology defines the physical structure of the crust [*Bassin et al.*, 2000] and geochemistry specifies the U, Th, and K abundances in continental crust [*Rudnick and Gao*, 2003] and oceanic crust [*White and Klein*, 2012]. Combining the seismological and geochemical information predicts 8 ±1 TW of radiogenic heating in the crust, which is consistent with an independent assessment based on heat flow [*Jaupart and Mareschal*, 2004]. Subtracting the radiogenic heating in the crust (≈8 TW) from the radiogenic heating either predicted by the geological models (11 – 38 TW) or suggested by estimates of planetary cooling (13 – 37 TW) poorly defines radiogenic heating in the mantle (3 – 30 TW). Cosmochemical models predict a low range of mantle radiogenic heating, spanning 3 – 13 TW, while geophysical models predict a high range, spanning 11 – 30 TW. Although the ranges of mantle radiogenic heating predicted by the two classes of geological models overlap slightly around 12

TW, the end-member values differ by an order of magnitude. Despite these large differences in predicted mantle radiogenic heating, there presently exist no experimental measurements that conclusively exclude either end-member prediction.

Intimately related to terrestrial radiogenic heating is a flux of electron antineutrinos, commonly called geo-neutrinos [*Fiorentini et al.*, 2007]. Beta decays of daughter nuclides in the radioactive series of $^{238}$U and $^{232}$Th produce detectable geo-neutrinos. Present detection techniques estimate the energy spectrum, but not the direction, of geo-neutrinos. Because the surface flux of geo-neutrinos hinges on proximity to radioactive sources, resolution of geo-neutrino signals from crust and mantle reservoirs typically requires knowledge of the local distribution of uranium and thorium. Geo-neutrinos are currently being detected underground in Japan [*Araki et al.*, 2005b; *Gando et al.*, 2011] and Italy [*Bellini et al.*, 2010]. Combining recent results from these projects constrains radiogenic heating to 15 – 41 TW, assuming the observations exhibit a specific thorium-to-uranium abundance ratio (Th/U=3.9) and sample the same homogeneous mantle signal. These encouraging results prompt an evaluation of the suitability of future observatories for further constraining the radioactive power of the planet.

This review discusses the present status and future prospects of measuring the radioactive power of the Earth by geo-neutrino observations. It begins with the physics of geo-neutrinos, opening with an introduction to neutrinos and proceeding with an evaluation of the antineutrino luminosity and radiogenic heating for uranium, thorium, and potassium (Section 2). Following this are descriptions of geo-neutrino interactions, including the current detection technique (Section 3) and the signal-distorting effects of neutrino oscillations (Section 4). It completes the physics input with a prescription for calculating geo-neutrino flux and signal (Section 5). Shifting to geology, this review constructs Earth models and calculates corresponding geo-neutrino signals, starting with very simple distributions of heat-producing elements (Section 6). Successive introductions of complexity, starting with the crust model (Section 7), refine Earth models and corresponding geo-neutrino signals (Section 8). A discussion of background to the geo-neutrino signal (Section 9) precedes a presentation of geo-neutrino observatories (Section 10). There follows an evaluation of recent geo-neutrino observations (Sections 11 and 12). A discussion of these results within the context of constraining global radioactive power, including possible future observations, ensues (Section 13). This review concludes with a presentation of the main results (Section 14).

## 2. Geo-neutrino production and energy spectra

Neutrinos and antineutrinos have associations with three electrically charged leptons, namely the electron, mu, and tau $(e, \mu, \tau)$. They are the lightest of the known massive particles and lack measureable electromagnetic properties [*Giunti and Studenikin*, 2009]. There are three known massive neutrino states $(\nu_1, \nu_2, \nu_3)$, which are linear combinations of the three known lepton flavor states $(\nu_e, \nu_\mu, \nu_\tau)$. Neutrinos travel as mass states and interact as flavor states, subject to flavor state oscillation in transit [*Fukugita and Yanagida*, 2003]. Because their dominant interaction with other particles is the weak nuclear force, they transit vast amounts of matter without deflection or appreciable absorption. Neutrinos and antineutrinos, which are produced by nuclear processes inside stars and planets, carry off otherwise-shrouded information about the interiors of these dense objects. Electron neutrinos emanate from the decay of neutron-poor nuclei, which are copiously produced by nuclear fusion inside stars. Electron antineutrinos emerge from the decay of neutron-rich nuclei, which are readily formed by nuclear fission reactors and naturally synthesized in supernovae. Detectable geo-neutrinos originate from terrestrial uranium and thorium, fused by an ancient stellar explosion, incorporated with the material forming the Earth, and distributed by global differentiation.

Terrestrial isotopes of uranium, thorium, and potassium contribute significantly to the planetary energy budget [*Van Schmus*, 1995]. These isotopes decay to stable nuclei with the release of energy and the emission geo-neutrinos according to:

$$^{238}U \rightarrow {}^{206}Pb + 8\alpha + 6e + 6\bar{\nu}_e + 51.698 \text{ MeV}$$
$$^{235}U \rightarrow {}^{207}Pb + 7\alpha + 4e + 4\bar{\nu}_e + 46.402 \text{ MeV}$$
$$^{232}Th \rightarrow {}^{208}Pb + 6\alpha + 4e + 4\bar{\nu}_e + 42.652 \text{ MeV}$$
$$^{40}K \rightarrow {}^{40}Ca + e + \bar{\nu}_e + 1.311 \text{ MeV } (89.3\%)$$
$$^{40}K + e \rightarrow {}^{40}Ar + \nu_e + 1.505 \text{ MeV } (10.7\%) \quad (1)$$

Calculation of the decay energy ($Q$) utilizes a standard table of atomic masses [*Audi and Wapstra*, 1995]. The average amount of terrestrial heat per decay is the decay energy minus the average energy carried off by geo-neutrinos, which originate almost exclusively in beta decay.

$$[A,Z] \rightarrow [A,Z+1] + e + \bar{\nu}_e + Q_\beta. \quad (2)$$

Conservation laws define the energy ($w_e$) and momentum ($p_e$) of the emitted electron. Using energy units ($mc^2 \rightarrow m; pc \rightarrow p$)

$$w_e = (Q_\beta + m_e - E_{\bar{\nu}_e}) \text{ and } p_e = \left[(Q_\beta + m_e - E_{\bar{\nu}_e})^2 - m_e^2\right]^{1/2}. \quad (3)$$

To close approximation the energy spectrum of antineutrinos from beta decay is proportional to

$$dn(E_{\bar{\nu}_e})/dE_{\bar{\nu}_e} \propto w_e E_{\bar{\nu}_e}^2 p_e^{2\gamma-1} e^{\pi\eta} |\Gamma(\gamma + i\eta)|^2, \quad (4)$$

where $\Gamma$ is the gamma function and *i* is the imaginary unit [*Preston*, 1962]. With $\alpha$ the fine structure constant

$$\gamma = \sqrt{1-\alpha^2(Z+1)^2} \text{ and } \eta = \alpha(Z+1)w_e/p_e. \tag{5}$$

Using the definitions in (3) gives the spectrum in terms of the antineutrino energy. The spectrum for each radioactive nuclide results from the sum of antineutrino spectra of all beta decays that lead to the stable daughter nucleus. These spectra contribute according to the branching ratios, fractional intensities, and endpoint energies of transitions to the various daughter nuclear states. Table 1 lists the symbols representing the quantities in this review.

Figure 1 displays the antineutrino intensity energy spectra per decay for $^{238}$U, $^{235}$U, $^{232}$Th, and $^{40}$K. The area under each spectrum equals the number of emitted antineutrinos per decay $n_{\bar{v}_e}$ of each parent nuclide. Multiplying this number by the average energy of each spectrum estimates the energy escaping the Earth on average each time a parent nuclide decays. The average energy of $^{40}$K requires a small correction, accounting for the emission of the neutrino from electron capture. Subtracting the average escape energy from the decay energy computes the radiogenic heat absorbed by the Earth on average per decay ($Q_h = Q - Q_v$). The calculated rate of heating per unit mass of the parent nuclide, or the isotopic heat generation, is

$$h = \frac{N_A \lambda}{\mu} Q_h, \tag{6}$$

with $N_A$ Avogadro's number, $\lambda$ the decay constant, and $\mu$ the molar mass. Table 2 presents the quantities used to calculate the radiogenic power of $^{238}$U, $^{235}$U, $^{232}$Th, and $^{40}$K. These quantities also allow calculation of the isotopic antineutrino luminosity

$$l = \frac{N_A \lambda}{\mu} n_{\bar{v}_e}. \tag{7}$$

Element specific heat generation and antineutrino luminosity follow from summing the isotopic values weighted by natural abundance. Table 3 presents values for uranium, thorium and potassium. These concur with values from similar recent calculations [*Enomoto et al.*, 2007; *Fiorentini et al.*, 2007]. Previous calculations of heat production tend to underestimate the contributions from uranium and thorium and overestimate the contribution from potassium at about the 4% level or less [*Hamza and Beck*, 1972; *Rybach*, 1988].

**3. Geo-neutrino detection**

Geo-neutrino detection presently exploits a coincidence of signals from quasi-elastic scattering on a free proton (hydrogen nucleus) in organic scintillating liquid. This follows the traditional method for real-time measurement of reactor antineutrinos, which was developed decades ago [*Reines and Cowan*, 1953]. In this neutron inverse beta decay reaction, an electron antineutrino becomes a positron by collecting the electric charge from a proton, which becomes a neutron [*Vogel and Beacom*, 1999].

$$\bar{\nu}_e + p \rightarrow e^+ + n \tag{8}$$

Both reaction products produce signals, correlated in position and time. The positron retains most of the available energy, which is approximately the electron antineutrino energy ($E_{\bar{\nu}_e}$) minus the difference between the rest mass energy of the neutron and proton ($\Delta = M_n - M_p$). It rapidly (<1 ns) loses kinetic energy through ionization, producing a prompt signal proportional to the energy of the electron antineutrino.

$$T_e = E_{\bar{\nu}_e} - \Delta - m_e \tag{9}$$

The positron soon annihilates with an electron, releasing $\gamma$-rays with total energy equal to twice the electron mass. If the $\gamma$-rays interact within the detector, typically by Compton scattering, this increases the energy and spatial spread of the prompt signal. Prior to annihilation, the positron has a significant probability (~50% in scintillating liquid) of briefly forming a bound state with an electron (positronium), delaying the annihilation signal by several nanoseconds [*Franco et al.*, 2011]. Although this delay degrades the positron position resolution, it provides a method for rejecting background.

The momentum of the electron antineutrino transfers principally to the neutron, initially moving forward and losing energy through collisions with hydrogen nuclei. Some of the recoiling protons contribute relatively small amounts of ionization energy to the prompt signal. After coming to thermal equilibrium, the neutron diffuses through the medium, typically for many microseconds before getting absorbed by an atomic nucleus within about a meter of the original interaction. Adding a neutron to the atomic nucleus triggers a release of radiation from the newly formed nuclide. Although reactor antineutrino experiments typically use gadolinium for neutron absorption, which decreases the diffusion time and significantly boosts the delayed signal [*Bemporad et al.*, 2002], present geo-neutrino observations detect the mono-energetic gamma ray from the formation of deuterium.

$$n + p \rightarrow d + \gamma \text{ (2.2 MeV)} \tag{10}$$

This delayed signal completes the temporal and spatial coincidence, confirming the detection of an electron antineutrino.

Detecting quasi-elastic proton scattering is an established method for measuring the energy spectrum of geo-neutrino interactions. The threshold energy is approximately the difference in mass energy between the neutron plus positron and the proton ($\Delta + m_e \approx$ 1.8 MeV). There is no sensitivity to geo-neutrinos with energy below this threshold, masking completely the contributions of $^{235}$U and $^{40}$K. Above the threshold energy, the cross section grows approximately as the square of the antineutrino energy minus the difference between the rest mass energy of the neutron and proton [*Vogel and Beacom*, 1999].

$$\sigma_p(E_{\bar{\nu}_e}) = 9.52(E_{\bar{\nu}_e} - \Delta)^2 \sqrt{1 - m_e^2/(E_{\bar{\nu}_e} - \Delta)^2} \times 10^{-44} \text{ cm}^2 \qquad (11)$$

The magnitude of this cross section (~$10^{-43}$ cm$^2$) coupled with the approximate geo-neutrino flux above the threshold energy (~$10^5$ cm$^{-2}$s$^{-1}$) requires a large number of free protons (~$10^{32}$) to realize a modest interaction rate (~10 y$^{-1}$). For hydrogen-rich liquids, such as mineral oil (CH$_2$) or water (H$_2$O), this corresponds to a detector of significant mass (~$10^6$ kg) and volume (~$10^3$ m$^3$). Geo-neutrino observatories, by the nature of weak interactions, are necessarily very large.

Currently operating geo-neutrino observatories, including those planned for near future operation, utilize organic scintillating liquids for the detection medium. These liquids typically contain mixtures of aromatic hydrocarbons, designed to respond quickly to ionizing radiation (several ns decay time) with relatively high photon production (~$10^4$ MeV$^{-1}$), and to have good optical transparency (>10 m attenuation length). Electrically charged particles traversing the scintillating liquid excite pi-bond electrons in benzene rings in the solvent to higher energy states. Relaxation of these excited states, typically through molecular collisions with fluorescing solutes, transfers the energy to photons. The emission spectra of the solutes match the transparency of the medium and the sensitivity of the light collectors. Light collection employs large (>20 cm diameter) hemispherical photomultiplier tubes, arrayed outside and looking into the scintillating liquid. Amplitude and timing information of the photomultiplier tube pulses resolves the position and energy of the ionizing particles. Resolutions, which determine geo-neutrino detection efficiency, improve with increased light collection.

Water is an inexpensive, optically transparent, hydrogen-rich liquid that could potentially serve as the detection medium for geo-neutrino observatories. In pure water light production by charged particles is limited to Cherenkov radiation, which is inherently weaker (~5%) than scintillation from organic liquids. However, much of the Cherenkov emission spectrum, which decreases as the inverse square of wavelength, is at wavelengths shorter than the sensitivity of the light collectors and the maximum transparency of water. Adding a fluorescing solute elevates the collected Cherenkov light by about a factor of three [*Dai et al.*, 2008; *Sweany et al.*, 2012]. Higher light levels are possible by dissolving organic scintillating liquid in water using surfactants (M. Yeh,

Water-based liquid scintillator for large-scale physics, presented at American Physical Society, Atlanta, April, 2012). The development of techniques for increasing collectable light levels while maintaining acceptable attenuation makes water a suitable, low cost detection medium for geo-neutrino observatories.

Quasi-elastic scattering of electron antineutrinos on nuclear targets other than hydrogen ($^1$H) potentially provides sensitivity to geo-neutrinos from $^{235}$U and $^{40}$K [*Krauss et al.*, 1984; *Kobayashi and Fukao*, 1991]. The nuclide with the lowest threshold energy, although probably not accessible in the quantity required for geo-neutrino observations, is $^3$He. A previously unidentified nuclear target with sensitivity to $^{235}$U and $^{40}$K geo-neutrinos is $^{106}$Cd (M. Chen, presented at Neutrino Sciences 2005, Honolulu, December, 2005), which is naturally abundant at the percent level. Cadmium tungstate, an inorganic scintillating crystal capable of particle identification by pulse-shape discrimination [*Fazini et al.*, 1998], is available for experimental applications (>200 g) with $^{106}$Cd enriched to the 66% level [*Belli et al.*, 2011]. A geo-neutrino interaction with $^{106}$Cd gives a double positron signature.

$$\bar{\nu}_e + {}^{106}Cd \rightarrow {}^{106}Ag + e^+; {}^{106}Ag \rightarrow {}^{106}Pd + e^+ + \bar{\nu}_e \text{ (24 min)} \qquad (12)$$

A clearly identifiable positron signature is of benefit here. Note, however, that the lifetime of positronium in inorganic crystals is perhaps shorter than in organic liquids. Moreover, the decay time of inorganic scintillating crystals is about a factor of 10 times those of organic scintillating liquids, making the signal from positronium formation [*Franco et al.*, 2011] possibly difficult to observe. This technique, although promising, requires further development.

A less developed method of detecting electron antineutrinos involves elastic scattering on atomic electrons [*Reines et al.*, 1976].

$$\bar{\nu}_e + e^- \rightarrow \bar{\nu}_e + e^- \qquad (13)$$

There is no energy threshold for this process, giving sensitivity to geo-neutrinos with energy below the neutron inverse beta decay threshold. The signal, however, is simply the recoiling electron. The lack of a coincidence tag and an intense flux of solar neutrinos make exploiting this reaction challenging. However, the electrons scatter in the forward direction, suggesting signal sensitivity through resolution of the electron direction. The maximum kinetic energy of the electrons is

$$T_{\max} = \frac{E_{\bar{\nu}_e}}{1 + m_e/2E_{\bar{\nu}_e}}. \qquad (14)$$

With the dependence on the weak mixing angle given by $x = 2\sin^2\theta_W$, the cross section is

$$\sigma_e(E_{\bar{\nu}_e}) = 0.43\left[x^2 T_{max} + (x+1)^2 \frac{E_{\bar{\nu}_e}}{3}\left\{1-(1-\frac{T_{max}}{E_{\bar{\nu}_e}})^3\right\} - x(x+1)\frac{m_e T_{max}^2}{2E_{\bar{\nu}_e}^2}\right] \times 10^{-44} \text{cm}^2.$$

(15)

Figure 2 displays the electron antineutrino cross sections over the range of energy relevant for geo-neutrino studies, showing both quasi-elastic scattering on protons and elastic scattering on electrons. For comparison, this figure includes the cross section of electron neutrino elastic scattering on electrons, which lacking interference between neutral and charged weak currents is larger by a factor of ~2.4 at these energies. This larger cross section exacerbates background from solar electron neutrinos.

The product of the antineutrino intensity energy spectra per decay of $^{238}$U, $^{235}$U, $^{232}$Th, and $^{40}$K (Fig. 1) and the scattering cross sections gives the geo-neutrino interaction energy spectra. Figure 3 displays the interaction energy spectra per decay of $^{238}$U, $^{235}$U, $^{232}$Th, and $^{40}$K for proton and electron scattering targets. Although electron scattering provides sensitivity to $^{40}$K and $^{235}$U and has an advantage of four electrons for each proton in organic scintillating liquid, finding the geo-neutrino signal in the intense background of solar neutrinos requires further development. Moreover, many of the scattered electrons have energy below the threshold for detection. Therefore, geo-neutrino detection by electron scattering is not considered further herein. This restricts sensitivity of geo-neutrino observations to the parent nuclides $^{238}$U and $^{232}$Th.

The number of antineutrinos per decay of the parent nuclide divided by the area under the interaction energy spectra calculates the flux of geo-neutrinos required to produce one interaction. The magnitude of this flux commands the exposure of geo-neutrino observations. It is convenient to scale this flux to one interaction during a fully efficient one-year exposure of $10^{32}$ protons, called a terrestrial neutrino unit (TNU) [*Mantovani et al.*, 2004]. This scaling provides the signal rate to flux conversion factors $(C = \varphi/R)$

$$C_U = 7.6\times 10^4 \text{ cm}^{-2}\text{ s}^{-1}\text{ TNU}^{-1} \text{ and } C_{Th} = 2.5\times 10^5 \text{ cm}^{-2}\text{ s}^{-1}\text{ TNU}^{-1}. \quad (16)$$

These factors, which agree with the results of previous calculations [*Enomoto et al.*, 2007; *Fiorentini et al.*, 2007], are sensitive to the shape of the beta decay spectrum [*Fiorentini et al.*, 2010].

**4. Geo-neutrino propagation**
Detectable geo-neutrinos travel at or near the speed of light in the vacuum [*Longo*, 1987], providing a near instantaneous view of terrestrial radioactivity. Neutrino oscillation diminishes the detectable geo-neutrino signal below inverse-square scaling, depending on energy and distance [*Araki et al.*, 2005a]. The survival probability of electron antineutrinos is

$$P_{ee}^{3\nu} = 1 - \{\cos^4(\theta_{13})\sin^2(2\theta_{12})\sin^2(\Delta_{21}) \\ + \sin^2(2\theta_{13})[\cos^2(2\theta_{12})\sin^2(\Delta_{31}) + \sin^2(2\theta_{12})\sin^2(\Delta_{32})]\},\tag{17}$$

where $\theta_{12}$ and $\theta_{13}$ are mixing angles, $\Delta_{ji} = 1.27(|\delta m_{ji}^2|L)/E_{\bar{\nu}_e}$ control the oscillations with $\delta m_{ji}^2 \equiv m_j^2 - m_i^2$ the neutrino mass-squared difference of $\nu_j$ and $\nu_i$ in eV$^2$, $L$ is the neutrino flight distance in meters, and $E_{\bar{\nu}_e}$ is the antineutrino energy in MeV. By definition the mass-squared differences satisfy the relation $\delta m_{31}^2 = \delta m_{32}^2 + \delta m_{21}^2$. An analysis of global neutrino data provides values for the mixing angles and the mass-squared differences [*Fogli et al.*, 2011]. A recent experimental measurement of the subdominant mixing angle $\theta_{13}$ [*An et al.*, 2012; *Ahn et al.*, 2012] establishes a non-zero value and reduces uncertainties. Noting that $\delta m_{31}^2 \approx \delta m_{32}^2 \gg \delta m_{21}^2$ suggests simplifying the survival probability to

$$P_{ee} \cong 1 - \left[\cos^4(\theta_{13})\sin^2(2\theta_{12})\sin^2(\Delta_{21}) + 0.5\sin^2(2\theta_{13})\right].\tag{18}$$

A 2 MeV geo-neutrino thus has an oscillation length of 65 +/-2 km, which is about 1% of the radius of the Earth. These determinations prompt an approximate expression for the average geo-neutrino survival probability, specifically

$$\langle P_{ee}\rangle \cong 1 - 0.5\left[\cos^4(\theta_{13})\sin^2(2\theta_{12}) + \sin^2(2\theta_{13})\right] = 0.544^{+.017}_{-.013}.\tag{19}$$

The new best value of the average survival probability is about 9% smaller than the ~0.58 value assumed by geo-neutrino studies prior to the discovery of a non-zero subdominant mixing angle [*Bellini et al.*, 2010; *Gando et al.*, 2011]. Uncertainty in the average survival probability as defined above (+3%/-2%), which is dominated by imprecision in the solar mixing angle, is small in comparison with the precision in the geo-neutrino flux (± 10 %) required to identify primordial heat loss.

**5. Terrestrial antineutrino signal**
Model estimates of the quantity and distribution of radioactive isotopes within the Earth allow calculation of the resulting geo-neutrino signal spectrum. The expression is

$$\frac{dN(E_{\bar{\nu}_e})}{dE_{\bar{\nu}_e}} = \varepsilon \frac{N_A \lambda}{\mu} \sigma_p(E_{\bar{\nu}_e}) \frac{dn(E_{\bar{\nu}_e})}{dE_{\bar{\nu}_e}} \langle P_{ee}\rangle \int_\oplus dV \frac{a(\vec{r}')\rho(\vec{r}')}{4\pi|\vec{r}-\vec{r}'|^2},\tag{20}$$

where $\varepsilon$ is the effective exposure (number free proton targets multiplied by observation time multiplied by detection efficiency) of a detector at position $\vec{r}$, $a(\vec{r}')$ is the parent nuclide abundance distribution, and $\rho(\vec{r}')$ is the Earth density distribution. In principle, the effective exposure involves an energy-dependent efficiency for detecting geo-neutrino interactions. The geo-neutrino signal spectrum sums the contributions from the parent nuclides $^{238}$U and $^{232}$Th. If the abundance of a parent nuclide is uniform throughout a spherical shell, then the integral

$$G = \int_{shell} dV \frac{\rho(\vec{r}')}{4\pi |\vec{r}-\vec{r}'|^2} \qquad (21)$$

calculates geological response factors [*Krauss et al.*, 1984]. Note that the denominator of the integrand is the square of the difference between two position vectors, making geo-neutrino flux calculation inherently different from Gauss' law.

A prediction of the geo-neutrino signal rate in TNU from a geological reservoir $G$ with uniform abundance of a parent nuclide $a$ is

$$R = \frac{\varphi}{C} = \frac{Gla}{C} \langle P_{ee} \rangle. \qquad (22)$$

A prediction of the elemental radiogenic heating in watts due to a geo-neutrino flux $\varphi$ from a geological reservoir $G$ with mass $M$ is

$$H = Mha = \frac{Mh\varphi}{Gl\langle P_{ee} \rangle} = \frac{MhRC}{Gl\langle P_{ee} \rangle}. \qquad (23)$$

If the mass ratio of thorium to uranium ($Th/U = a_{Th}/a_U$) is uniform throughout the reservoir, then

$$\phi_{Th}/\phi_U = 0.217 Th/U \text{ and } R_{Th}/R_U = 0.066 Th/U. \qquad (24)$$

These relationships are useful for interpreting existing geo-neutrino observations. The observed interaction energy spectrum at a given observatory depends on the relative fluxes from thorium and uranium. Figure 4 shows spectra for different values of Th/U.

Given a measured geo-neutrino signal, application of the average neutrino survival probability overestimates the abundances of parent nuclides and underestimates the thorium-to-uranium ratio. Figure 5 shows the ratio of $P_{ee}$ to <$P_{ee}$>, which is greater than unity and rising with energy for simulated continental and oceanic observations. Sources of uranium and thorium within one oscillation length of the observatory are the cause, making the effect more pronounced on continental crust than oceanic crust. Although these effects at the level of several percent are small in comparison with present experimental uncertainties, they warrant consideration in future measurements, especially at continental locations.

## 6. Simple Earth models

Previous sections present the physics input required for calculating radiogenic heating and surface geo-neutrino signal. Geology guides the remaining input. The global distribution of radiogenic heating significantly influences the structure, dynamic activity, and thermal history of the planet [*Jaupart et al.*, 2007; *Korenaga*, 2008; *Lay et al.*, 2008]. This distribution also influences the surface geo-neutrino flux [*Mareschal et al.*, 2011]. This section constructs simple model predictions of the geo-neutrino signal.

Seismology resolves a stratified shell-structure of the planet. A preliminary reference Earth model (PREM) identifies a solid inner and a liquid outer core, a lower and an upper mantle, and a uniform crust [*Dziewonski and Anderson*, 1981]. This model specifies thickness and density of concentric spherical shells, facilitating calculation of geological response factors [*Krauss et al.*, 1984]. Table 4 presents relevant geophysical information, including response factors and masses of the various sub-shells and the entire Earth. Cosmochemistry provides information on U and Th abundances in primitive chondritic meteorites [*Wasson and Kallemeyn*, 1988], which provide a convenient reference for comparing models. The synthetic Earth models presented in this review assume chondritic abundances of refractory U (7.8 ppb ±10 %) and Th (29.8 ppb ±10 %) [*Palme and O'Neill*, 2003]. The abundance of moderately volatile K follows from a terrestrial K/U ratio of 12,000 [*Arevalo et al.*, 2009]. Figure 6 compares the reference U and Th abundances with other estimates [*Rocholl and Jochum*, 1993; *McDonough and Sun*, 1995].

Very simple Earth models estimate the terrestrial antineutrino signal from concentrations of uranium and thorium uniformly distributed throughout the whole planet. These radioactively undifferentiated Earth models assume reference abundances of U, Th, and K. Distributing these heat-producing elements uniformly throughout Earth layers as defined by PREM produces radiogenic heating of ≈11 TW. This simple cosmochemical Earth model, providing a lower bound on radiogenic heating, estimates a geo-neutrino signal of ≈9 TNU. Increasing the reference abundances of U, Th, and K by approximately a factor of four produces radiogenic heating equivalent to the surface heat flow. This simple fully radiogenic Earth model estimates a geo-neutrino signal of ≈35 TNU.

An initial refinement of these very simple models recognizes geological evidence for a metallic iron-nickel planetary core, virtually devoid of uranium, thorium, and potassium [*McDonough*, 2003]. The process of core formation, effecting planetary scale geochemical differentiation, clearly does not alter the amount of radiogenic heating. However, it does increase reference U, Th, and K abundances by a factor ≈1.5 in the non-metallic, silicate outer shell, or primitive mantle (PM). This bulk silicate Earth enrichment of uranium and thorium by core formation increases geo-neutrino signals by ≈11 % relative to estimates for undifferentiated Earth models. These semi-differentiated cosmochemical and fully radiogenic Earth models estimate geo-neutrino signals of ≈10 and ≈39 TNU, respectively.

**7. Crust model**
Further refined Earth models incorporate knowledge of the physical and chemical heterogeneity of the crust of the planet. A seismological model defines the thicknesses and densities of crust layers and sediments on a 2-degree by 2-degree grid [*Bassin et al.*, 2000]. This model describes continental crust covering 40.5% of the surface with an average density of 2.9 g/cm$^3$ and an average thickness of 34.3 km. The complementary 59.5% of the surface area is covered by oceanic crust with average density 3.6 g/cm$^3$ and an average thickness of 5.8 km. Separate geochemical models estimate the abundances of uranium, thorium, and potassium in bulk oceanic crust [*White and Klein*, 2012] and in the upper, middle, and lower layers of continental crust [*Rudnick and Gao*, 2003]. These geochemical models specify abundances of U, Th, and K in bulk oceanic crust a factor of 20 to 30 less than in bulk continental crust. The crust model assigns upper continental crust abundances of U, Th, and K to all sediments. Combining the physical and geochemical crust models assesses the total masses of uranium, thorium, and potassium in the bulk crust, estimating 8.1 ±0.8 TW of radiogenic heating. This value is consistent with an independent estimate of radiogenic heating in continental crust derived from heat flow [*Jaupart and Mareschal, 2004*]. Table 5 summarizes the relevant crust model information.

The crust model describes average continental crust with about two orders of magnitude more U, Th, and K per unit area than average oceanic crust. This heterogeneity introduces large spatial variations in the terrestrial antineutrino signal. A spherical shell of average continental crust would produce a geo-neutrino signal of 34 ±5 TNU, while a spherical shell of average oceanic crust would produce a geo-neutrino signal of 0.4 ±0.1 TNU. This variability of the geo-neutrino signal from the crust is an important consideration for locating a geo-neutrino observatory. According to the crust model, the maximum geo-neutrino signal from the crust, which occurs near the Himalayas, is 52 ±7 TNU. The minimum geo-neutrino signal from the crust, which occurs near the equator north of the Tuamotu Islands in Oceania, is 3.2 ±0.4 TNU. A typical geo-neutrino signal from the crust at a continental location is 40 ±6 TNU. The abundance of uranium, which accounts for almost 80% of the crust signal in this model, dominates the uncertainty.

**8. Refined Earth models**
The construction of refined Earth models employs the crust model and assumes the core is free of the main heat-producing elements. With these elements in approximately fixed amounts in the crust, the complementary amounts in the depleted mantle (DM) become the model discriminant. The composition of the source reservoir of mid-ocean ridge basalts (MORB), or depleted MORB mantle (DMM), provides a lower bound on possible abundances of U, Th, and K in the depleted mantle. Table 6 summarizes the estimates of heat-producing element abundances in depleted MORB source mantle.

This section presents the mantle radiogenic heating and geo-neutrino signals predicted by the end members of the cosmochemical (CC) and geophysical (GP) models.

Modeling the abundances of U and Th in the bulk silicate Earth enriched by ≈1.5 relative to chondritic values, and setting K/U at 12,000, is consistent with an enstatite Earth [*Javoy et al.*, 2010]. This low end-member cosmochemical model provides mantle radiogenic heating of ≈3.2 TW. This predicts global radiogenic heating of ≈11 TW, mitigating the apparent missing atmospheric argon problem [*Allègre*, 1996]. A homogeneous distribution of these heat-producing elements in the depleted mantle yields abundances consistent with several composition estimates for depleted MORB source mantle [*Jochum et al.,* 1983; *Salters and Stracke*, 2004; *Workman and Hart,* 2005; *Boyet and Carlson,* 2006], yet lower than another estimate [*Arevalo and McDonough*, 2010]. Cosmochemical models typically specify U and Th in the bulk silicate Earth enriched by 2.5 – 2.8 relative to chondritic abundances [*Hart and Zindler*, 1986; *McDonough and Sun*, 1995; *Palme and O'Neill*, 2003], significantly more than expected from core segregation alone. However, a new statistical analysis recommends intermediate values of 1.8 – 2.5 [*Lyubetskaya and Korenaga*, 2007]. The high end-member enrichment value of 2.8 predicts mantle radiogenic heating of ≈13 TW. As a general class, cosmochemical models allow mantle radiogenic heating of 3.2 – 13 TW.

Depending on the distribution of heat-producing elements in the mantle, more heating correlates with higher geo-neutrino signal. Assuming abundances do not decrease with depth, a homogeneous distribution, which associates with whole mantle convection, produces the largest geo-neutrino signal. Seismic images of slabs of oceanic lithosphere sinking into the deep mantle support whole mantle convection [*van der Hilst et al.,* 1997]. If enrichment is <2, mass balance tolerates whole mantle convection of primitive mantle with DMM composition, depending somewhat on the DMM estimate. Figure 7a plots the allowed mass of depleted mantle with DMM composition relative to present-day mantle mass as a function of U and Th enrichment. Enrichment >2, including the standard cosmochemical model values of 2.5 – 2.8, allow layered convection with depleted MORB source mantle overlying isolated primitive mantle. According to mass balance, primitive mantle comprises at least 20% of the present-day mantle for enrichment <5. Despite the images of sinking slabs, the possibility for a deep untapped primitive mantle reservoir remains [*Allègre*, 1997]. Indeed, recent chemical isotope analyses favor the survival of isolated deep mantle reservoirs [*Boyet and Carlson*, 2005; *Jackson et al.*, 2010]. For a given enrichment, distributions of U and Th abundances that increase with depth decrease the mantle geo-neutrino signal. A lower bound on the signal results from maximizing U and Th abundances in a relatively thin basement layer, for example D", beneath depleted mantle with DMM composition. This distribution allows whole mantle convection and an isolated geochemical reservoir [*Tolstikhin and Hofmann*, 2005]. Figure 7b plots the abundances of the heat-producing elements in the base layer relative to primitive mantle as a function of enrichment. These different heat-producing element distributions have significant geological implications. At the lowest enrichment of ≈1.5 the geo-neutrino signal from the D" distribution is ≈5 % less than the signal from the homogeneous distribution. The signal deficit grows with enrichment, reaching ≈20% at enrichment of 2.8. The geo-neutrino signals predicted by the cosmochemical model low and high end-members, representing enrichments of 1.5 and 2.8, for the homogeneous distribution are 2.7 and 11 TNU, respectively.

Physical Earth models recognize mantle convection as the main mechanism for the transport of interior heat to the surface. Sufficient convection requires viscosity consistent with surface heat flow. The standard approach uses a scaling law to relate heat flow and viscosity, predicting the thermal evolution of the Earth. Refining the description of convection, mainly focusing on mantle viscosity, and tuning the strength of this relationship selects solutions consistent with estimates of previous and present conditions. The convective Urey ratio [*Korenaga,* 2008] with heat flow out of the mantle ≈36 TW normalizes the comparison of geophysical models. If temperature alone influences viscosity, then the required radiogenic heating in the buk silicate Earth is as high as ≈38 TW, which is equivalent to a convective Urey ratio ≈0.8 [*Turcotte,* 1980]. Including the effects of water on viscosity, however, reduces silicate Earth heating requirement to ≈19 TW, which is equivalent to a convective Urey ratio ≈0.3 [*Crowley et al.,* 2011]. These radiogenic heating requirements provide upper and lower limits for the geophysical model. An intermediate radiogenic heating requirement results from considering variable viscosity convection [*Christensen,* 1985]. This suggests a convective Urey ratio of ≈0.5, corresponding to primitive mantle radiogenic heating of ≈27 TW. The geophysical model (GP), covering convective Urey ratios of 0.3 – 0.8, specifies mantle radiogenic heating of 11 – 30 TW. The geo-neutrino signal from the homogeneous mantle distribution is 9.5 – 26 TNU, while the signal from the enriched D" distribution is ≈20 % less.

The cosmochemical (CC) and geophysical (GP) models predict 11 – 21 TW and 19 – 38 TW of radiogenic heating in the bulk silicate Earth, respectively. Table 7 summarizes these amounts and other relevant information for these models. Subtracting radiogenic heating in the bulk crust of ≈8 TW from these model values for the silicate Earth results in predictions of 3 – 13 TW and 11 – 30 TW of radiogenic heating in the depleted mantle for the cosmochemical and the geophysical models, respectively. The differences in mantle heating correlate with dissimilar mantle geo-neutrino signals. Assuming homogeneous distribution of U and Th in the depleted mantle predicts geo-neutrino signals of 3.2 – 11 TNU and 9.5 – 26 TNU for the cosmochemical and geophysical models, respectively. For comparison, the refined fully radiogenic Earth model, allowing for no primordial heat loss, predicts 35 – 41 TW of mantle radiogenic heating and a mantle geo-neutrino signal of 29 – 34 TNU. Table 8 summarizes the relevant depleted mantle information. Resolving mantle geo-neutrino signals to constrain geological models requires an understanding of background.

**9. Background**
Identified background to the geo-neutrino signal takes several forms. It comes from electron antineutrinos from sources other than the Earth and from forms of radiation that mimic the quasi-elastic scattering coincidence. This section describes the identified background and discusses methods for reduction.

A nuclear reactor emits an intense and well-studied flux of electron antineutrinos [*Mueller et al.*, 2011]. This isotropic flux is proportional to the reactor power. Nuclear reactors in operation worldwide, which have thermal power exceeding 1 TW, contribute some level of electron antineutrino flux in any observatory. This flux initiates quasi-elastic scattering coincidences, which are indistinguishable from geo-neutrino interactions. Because shielding is not possible, reduction requires distancing the observatory from reactors. In practice, the locations and power history of reactors determines the expected contribution to the background, allowing subtraction with reasonable accuracy (~5%). A crude estimate of background from nuclear reactors assumes 1 TW of power at a distance of one Earth radius, which produces a rate of ≈2 TNU in the geo-neutrino energy window. Although sites can experience rates several orders of magnitude higher, this estimates the minimum expected rate of background from nuclear reactors.

Known natural sources of electron antineutrinos, other than the Earth, include supernovae and the atmosphere. The gravitational collapse of a stellar core produces a hot proto-neutron star, which cools primarily through the emission of neutrino-antineutrino pairs. Although the energy of these neutrinos is typically several tens of MeV at the time of emission, this energy degrades as the universe expands over time. Reasonable assumptions for the rate of stellar formation this expansion predict a diffuse flux of electron antineutrinos from all past supernovae is ~10 $cm^{-2}s^{-1}$ in the geo-neutrino energy window [*Ando*, 2004]. Primary cosmic rays bombard the upper atmosphere, creating showers of unstable pions and muons. The decay of these short-lived particles produces a flux of atmospheric neutrinos. The predicted flux of atmospheric antielectron neutrinos with energy in the geo-neutrino window is less than 1 $cm^{-2}s^{-1}$ [*Gaisser et al.*, 1988]. The fluxes of supernovae relic neutrinos and atmospheric neutrinos are very small compared with the ~$10^5$ $cm^{-2}s^{-1}$ flux of geo-neutrinos. Background from these natural sources of electron antineutrinos is negligible.

Muons from primary cosmic ray interactions in the upper atmosphere penetrate the planet to great depths [*Crouch et al.*, 1978], causing background in geo-neutrino observatories. Although an energetic muon loses energy (~2 MeV/cm) primarily through interactions with atomic electrons, it is the occasional collision with an atomic nucleus that produces background to the geo-neutrino signal. These collisions can initiate nuclear reactions, producing energetic neutrons [*Mei and Hime*, 2006] and neutron rich isotopes [*Abe et al.*, 2010]. Energetic neutrons that originate inside the detector, or enter from outside, lose energy through collisions with protons in the scintillating liquid. The deposited ionization energy produces scintillation, which can mimic the prompt signal. The neutron upon capture can produce the delayed signal, completing the coincidence. An estimate of background from fast neutrons is <0.8 TNU [*Bellini et al.*, 2010]. Two neutron-rich isotopes, specifically $^8$He and $^9$Li, have significant branching fractions for decay by emission of a neutron and a beta minus. The beta minus produces the prompt signal and the neutron upon capture produces the delayed signal, mimicking the quasi-elastic scattering coincidence. Background from these isotopes is approximately 0.5 TNU [*Araki et al.*, 2005b; *Bellini et al., 2010*]. Reduction requires an overburden of earth or water, usually measured in meters of water equivalent (m.w.e), to attenuate the muon flux.

One in ninety naturally occurring carbon atoms is $^{13}$C, which captures alpha particles. One identified alpha producing detector contaminant is $^{210}$Po, which is in the decay series of $^{238}$U and makes alpha particles with 5.3 MeV of kinetic energy. The capture of alpha particles by $^{13}$C produces a neutron with up to 7.3 MeV of kinetic energy. The reaction is written as $^{13}$C($\alpha$,n)$^{16}$O. The fast neutron mimics the quasi-elastic scattering coincidence as explained above. This is a dominant source of background if there is radon contamination throughout detector [*Araki et al.*, 2005b]. Purification reduces this background to <0.3 TNU [*Bellini et al.*, 2010].

Radio-purity of detector components and shielding around the detector reduce accidental background. This background contribution comes mostly from gamma rays from trace-level concentrations of uranium and thorium in detector components and matter surrounding detector. Estimates of this background are 1.3+/-0.2 TNU [*Bellini et al.*, 2010] and 3.4+0.2 TNU [*Araki et al.*, 2005b].

The most serious background to the geo-neutrino signal comes from antineutrinos from nuclear reactors. Antineutrinos from the atmosphere and from all past supernovae are negligible. The antineutrino background rate from a 6 GW nuclear reactor in an observatory located ~100 km distant is ≈50 TNU, which is comparable to predicted geo-neutrino signal at a continental location. Reactor background of this level introduces systematic uncertainty in the geo-neutrino signal rate of <3 TNU after subtraction. Radio-purity of detector components, shielding, and overburden are necessary to control avoidable (non-antineutrino) background to <3 TNU. This low rate, which is estimated with ≈15 % uncertainty, is consistent with that achieved at an observatory with an overburden ≈3 km.w.e [*Bellini et al.*, 2010]. This introduces systematic uncertainty in the geo-neutrino signal rate ≈0.4 TNU after subtraction.

**10. Geo-neutrino observatories**
Geo-neutrino observatories are currently operating at two northern hemisphere locations. These observatories, one in Japan [*Gando et al.*, 2011] and one in Italy [*Bellini et al.*, 2010], monitor large volumes of organic scintillating liquids for the delayed coincidence signal, indicative of electron antineutrino quasi-elastic scattering on protons.

The Kamioka Liquid-Scintillator Antineutrino Detector (KamLAND) sits under Mt. Ikenoyama (36.42° N, 137.31° E) in Japan [*Gando et al.*, 2011], providing an equivalent flat overburden of 2.05 ±.15 km.w.e. [*Mei and Hime*, 2006]. It monitors ~6x10$^{31}$ free protons with >1800 photomultiplier tubes, collecting scintillation light from 34 % of solid angle. The detector selects geo-neutrino signals from uranium and thorium with efficiencies of 80.7 % and 75.1 %, respectively [*Gando et al.*, 2011]. These efficiencies account for hardware and software selection criteria.

The Borexino detector sits under Mt. Aquila (42.45° N, 13.57° E) in Italy [*Bellini et al.*, 2010], providing an equivalent flat overburden of 3.1 ±0.2 km.w.e. [*Mei and Hime*, 2006]. It monitors ~1x10$^{31}$ free protons with ~2200 photomultiplier tubes, collecting scintillation light from ~30 % of solid angle. The detector selects geo-neutrino signals from uranium and thorium with efficiency of 85±1 % [*Bellini et al.*, 2010].

Although the locations of the operating geo-neutrino observatories are not optimal for constraining the thermal evolution of the Earth, they do remotely monitor terrestrial radiogenic heating with significant measurements of the geo-neutrino flux. These measurements presently carry precision of about 40 %. The following presents a uniform analysis of these results, leading to estimates of the radiogenic heat production and comparisons with the selected Earth models.

**11. Japan measurement**

The KamLAND collaboration reports an observation of 841 ±29 candidate antineutrino events during a 3.49 TNU$^{-1}$ detector exposure [*Gando et al.*, 2011]. They estimate 730 ±32 background events, primarily from nearby nuclear reactors. This leads to a background-subtracted geo-neutrino signal of 111 +45/-43 events, which excludes the null hypothesis at the 99.5 % confidence level. Note that uncertainty in the predicted background (±32) is greater than the statistical error on the total number of observed events (±29), indicating that systematic uncertainty dominates the precision of the KamLAND geo-neutrino signal. After correcting for detection efficiency, the background-subtracted geo-neutrino signal rate is 40.0 ±10.5 (stat) ±11.5 (sys) TNU.

An unconstrained maximum likelihood analysis of the KamLAND data prefers 65 events from uranium and 33 events from thorium [*Gando et al.*, 2011]. Using the conversion factors in eq. (16) this corresponds to a flux (signal rate) of 1.75 cm$^{-2}$μs$^{-1}$ (23.1 TNU) from uranium and 3.15 cm$^{-2}$μs$^{-1}$ (12.6 TNU) from thorium. By eq. (24) this division of events implies a signal-averaged thorium-to-uranium mass ratio of 8.3, which is more than twice the chondritic ratio. However, the one standard deviation uncertainty contour allows no events from either uranium or thorium, corresponding to a signal averaged thorium-to-uranium mass ratio of either infinity or zero, respectively. Deriving geological conclusions from the KamLAND data requires assumptions about the distribution of uranium and thorium in the geo-neutrino source regions.

Assuming the KamLAND data result from a signal-averaged thorium-to-uranium mass ratio of 3.9 estimates 4.3 +1.2/-1.1 cm$^{-2}$μs$^{-1}$ of flux or a signal rate of 38 ±10 TNU [*Gando et al.*, 2011]. This cosmic mass ratio specifies 2.3 ±0.6 cm$^{-2}$μs$^{-1}$ of flux (30 ±8 TNU) from uranium and 2.0 ±0.5 cm$^{-2}$μs$^{-1}$ of flux (8 ±2 TNU) from thorium. A detailed study of the crust, including variations in the local geology, predicts 1.59 ±0.25 cm$^{-2}$μs$^{-1}$ of flux (20.9 ±3.3 TNU) from uranium and 1.34 ±0.21 cm$^{-2}$μs$^{-1}$ of flux (5.4 ±0.9 TNU) from thorium [*Enomoto et al.*, 2007]. Guidance from another study assigns the 16% uncertainty [*Coltorti et al.*, 2011]. The difference between the observed flux and the predicted crust flux estimates 1.3 +1.3/-1.2 cm$^{-2}$μs$^{-1}$ of mantle flux or a signal rate of 12 ±11 TNU.

The following presents assessments of radiogenic heating, resulting from two distributions of uranium, thorium, and potassium in the mantle, and which are consistent with the mantle flux estimated by KamLAND data. Distributing these elements homogeneously sets the minimum value of 14 +14/-13 TW of radiogenic heating in the mantle. A maximum value results from assuming an enriched 150-km thick layer (geological response factor for D" in Table 4) of heat-producing elements ($a$(U)≈0.28 ppm; $a$(Th)≈1.9 ppm) at the base of a homogeneous mantle with radiogenic composition estimated by the source material of mid-ocean ridge basalts [*Salters and Stracke*, 2004]. This distribution leads to 18 ±18 TW of radiogenic heating in the mantle. Adding the crust contribution of 8 ±1 TW to the minimum and maximum mantle heating values provides global assessments of 22 +14/-13 TW and 26 ±18 TW of terrestrial radiogenic heating, respectively. Comparing these results with the 43 - 49 TW surface heat flow suggests the presence of primordial heat loss if the distribution of U and Th in the mantle is homogeneous.

**12. Italy measurement**

The Borexino collaboration reports an observation of 15 candidate antineutrino events from a 0.179 TNU$^{-1}$ detector exposure [*Bellini et al.*, 2010]. They estimate 5.0 ±0.3 events from nearby nuclear reactors and 0.31 ±0.05 events from non-antineutrino sources. This leads to a background-subtracted geo-neutrino signal of 9.7 ±3.9 events, which excludes the null hypothesis at the 99.4 % confidence level. Note that statistical error on the total number of observed events (±3.9) is much greater than the uncertainty in the predicted background (±0.3). The reported systematic error demonstrates the potential for measuring the geo-neutrino signal with an ultimate precision of less than 10 %. This is due in part to the radio-purity of the Borexino detector. After correcting for detection efficiency, the background-subtracted geo-neutrino signal rate is 64 ±25 (stat) ±2 (sys) TNU.

An unconstrained maximum-likelihood analysis of the Borexino data prefers 9.9 +4.1/-3.4 events [*Bellini et al*., 2010]. As with the KamLAND data, the present precision of the Borexino data prevents geological conclusions without assumptions about the distribution of uranium and thorium in the source regions. Assuming the Borexino data result from a signal-averaged thorium-to-uranium mass ratio of 3.9 estimates 7.3 +3.0/-2.5 cm$^{-2}$μs$^{-1}$ of flux or a signal rate of 65 +27/-22 TNU. This cosmic mass ratio specifies 3.9 +1.6/-1.4 cm$^{-2}$μs$^{-1}$ of flux (52 +21/-18 TNU) from uranium and 3.3 +1.4/-1.1 cm$^{-2}$μs$^{-1}$ of flux (13 +6/-5 TNU) from thorium. A detailed study of the crust, including variations in the local geology, predicts 1.50 ±0.25 cm$^{-2}$μs$^{-1}$ of flux (19.7 ±3.3 TNU) from uranium and 1.36 ±0.22 cm$^{-2}$μs$^{-1}$ of flux (5.4 ±0.9 TNU) from thorium [*Coltorti et al*., 2011]. The difference between the observed flux and the predicted crust flux estimates 4.4 +3.0/-2.5 $cm^{-2}\mu s^{-1}$ of mantle flux or a signal rate of 40 +27/-23 TNU.

The following presents assessments of radiogenic heating, resulting from two distributions of uranium, thorium, and potassium in the mantle, and which are consistent with the mantle flux estimated by Borexino data. Distributing these elements homogeneously sets the minimum value of 48 +33/-27 TW of radiogenic heating in the mantle. A maximum value results from assuming an enriched 150-km thick layer (geological response factor for D" in Table 4) of heat-producing elements ($a$(U)≈1.8 ppm; $a$(Th)≈6.9 ppm) at the base of a homogeneous mantle with radiogenic composition estimated by the source material of mid-ocean ridge basalts [*Salters and Stracke*, 2004]. This distribution leads to 62 +44/-36 TW of radiogenic heating in the mantle. Adding the crust contribution of 8 ±1 TW to the minimum and maximum mantle heating values provides global assessments of 56 +33/-27 TW and 70 +44/-36 TW of terrestrial radiogenic heating, respectively. These results are consistent with 43 – 49 TW surface heat flow, providing no evidence for primordial heat loss.

**13. Discussion**
Two observations of the surface geo-neutrino signal by detectors separated by about 120° of longitude in the northern hemisphere now exist. Systematic uncertainty of the background limits the precision of the KamLAND observation, while low statistics limits the precision of the Borexino observation. The resulting precisions of about 40 % do not allow measurement of the thorium-to-uranium mass ratio. Assuming the observations exhibit a specific thorium-to-uranium mass ratio (Th/U=3.9), allows assessments of terrestrial radiogenic heating.

In general, higher geo-neutrino signal rates indicate higher levels of radiogenic heating. However, the distribution of U and Th within the mantle influences the geo-neutrino signal. The maximum signal results from a homogeneous distribution, assuming abundances of U and Th do not decrease with depth in the mantle. The minimum signal places as much U and Th in an enriched layer at the base of the mantle (D") as allowed by a depleted mantle with depleted MORB source mantle composition. Figure 8a compares the maximum and minimum mantle geo-neutrino signals with the combined observations of KamLAND and Borexino as a function of U and Th enrichment. This comparison demonstrates the potential to exclude enrichment parameters using measurements of the mantle geo-neutrino signal rate. Additional leverage results from measurements of the mantle geo-neutrino signal Th/U ratio. This follows from the super-chondritic Th/U ratio (Th/U ≈4.3) in continental crust, forcing a sub-chondritic Th/U ratio in the depleted mantle. Figure 8b shows the dependence of the Th/U signal ratio on U and Th enrichment bound by the mantle distributions for maximum and minimum signal rate.

The preceding discussion describes how the radiogenic heating indicated by a geo-neutrino signal depends on the assumed distribution of U and Th within the mantle.

Figure 9 compares the minimum and maximum radiogenic heating assessments of KamLAND and Borexino with the predictions of the geological models and with the surface heat flow. The KamLAND result assuming a homogenous distribution of U and Th in the mantle (minimum radiogenic heating), suggesting the presence of primordial heat loss, mildly excludes the synthetic fully radiogenic model and the high end-member of the geophysical model. These exclusions do not hold for the KamLAND result assuming an enriched layer at the base of the mantle (maximum radiogenic heating). The Borexino result (both maximum and minimum), implying radiogenic heating in the silicate Earth > 29 TW, excludes the presence of primordial heat loss, the cosmochemical model, and the low end-member of the geophysical model. However, the Borexino result, which is based on relatively small statistics, carries large uncertainties. Assuming both the KamLAND and Borexino observatories sample identical mantle signals allows combining the two results in a weighted average. The combined result, assuming a homogeneous mantle, mildly excludes the synthetic fully radiogenic model and the low end-member of the cosmochemical model. Assuming an enriched layer retains the mild exclusion of the low end-member of the cosmochemical model and loses the exclusion of the synthetic fully radiogenic model. If the two observatories sample different mantle signals, it is tempting to speculate that the higher Borexino value is due in part to proximity to the large low shear velocity province beneath the African continent [*Garnero and McNamara*, 2008] with enriched concentrations of heat-producing elements.

The successful measurements of geo-neutrinos by KamLAND and Borexino prompt an evaluation of radiogenic heating measurements from future observatories. Of near term interest is the SNO+ project, which is deploying a detector in a mine (46.47° N, 278.80° E) near Sudbury, Canada at a depth of ~6 km.w.e. [*Chen*, 2006]. It plans to monitor ~$6 \times 10^{31}$ free protons with >9000 photomultiplier tubes, collecting scintillation light from 54 % of solid angle. Figure 10 shows the energy spectrum of antineutrino interactions predicted at the Sudbury site, assuming a geo-neutrino rate of 49 TNU with a signal averaged Th/U ratio of 4.1 and a reactor background rate of 40 TNU in the geo-neutrino energy region. After an exposure of 3 $TNU^{-1}$, these rates, assuming 10 TNU from the mantle, 15 % uncertainty in the geo-neutrino rate from the crust, and ignoring the negligible non-neutrino background, project a 12 % measurement of the geo-neutrino rate and ≈80 % measurement of the mantle geo-neutrino rate [*Dye*, 2010]. The latter corresponds to estimates of global and mantle radiogenic heating equal to 20 ±10 TW and 12 ±10 TW, respectively. Increasing exposure gains little, as the ±10 TW uncertainty is not much larger than the ±8 TW systematic limit for the given assumptions. Another project, LENA, plans to deploy a much larger detector, either in a mine (63.66° N, 26.05° E) near Pyhäsalmi, Finland or under the French-Italian Alps (45.14° N, 6.69° E) near Modane, France [*Wurm et al.*, 2012]. The high statistics possible with this project allows measurement of the signal averaged thorium-to-uranium mass ratio to 20 % or better [*Wurm et al.*, 2012]. Unfortunately, model predictions of the signal averaged thorium-to-uranium ratio at the existing and the future continental locations have relatively small differences (<3 %). Moreover, the rate shift and spectral distortion introduced by assuming the average neutrino survival probability (Fig. 5) overestimates mantle signal and underestimates the Th/U ratio. These effects are enhanced at locations with proximity to increased levels of uranium and thorium, such as the Sudbury basin [*Perry et al.*, 2009].

Deploying a detector in the ocean basin far from continental crust allows maximal sensitivity to the geo-neutrino signal from the mantle [*Rothschild et al.*, 1998; *Enomoto, et al.*, 2007; *Dye,* 2010; *Gando et al.*, 2011; *Mareschal et al.*, 2011]. The predicted signal from the crust over much of the Pacific is less than 4 TNU [*Enomoto et al.*, 2007]. This low crust rate reduces the systematic uncertainty of the mantle signal to ≈1 TNU, which is comparable to that introduced by the ±10 % precision of the chondritic abundances of uranium and thorium. Figure 11 plots the expected total geo-neutrino signals with systematic uncertainty only, assuming a homogeneous mantle, for an oceanic, existing, and continental site as a function of radiogenic heating. Assuming radiogenic heating measurements of 20 TW, which is the overlap of the cosmochemical and geophysical model predictions, estimates the ultimate precisions from a single observation. The oceanic, existing, and continental observations expect to measure 20 TW of radiogenic heating with uncertainty no better than ≈15 %, ≈30 %, and ≈40 %, respectively. Clearly, the oceanic observation offers the best resolution of radiogenic heating and the tightest constraints on geological models. Moreover, the oceanic observation potentially offers discrimination by geo-neutrino signal thorium-to-uranium ratio. The cosmochemical and geophysical models predict signal averaged thorium-to-uranium ratio ranges at the oceanic observatory of 3.2 – 3.6, and 3.5 – 3.7, respectively. Distinguishing these different values requires analysis of the shape of the interaction energy spectrum (see Fig. 4), which appears possible with sufficient exposure [*Wurm et al.*, 2012]. Discriminating the geo-neutrino signal rates predicted by the geological models at the oceanic site becomes a statistical issue of detector exposure. Constraining the model end-members requires a relatively modest exposure of about 2 TNU$^{-1}$. Error in the mantle geo-neutrino measurement remains dominated by statistics for exposures of 20–50 TNU$^{-1}$, depending inversely on the level of radiogenic heating. A project to build and operate a movable deep ocean antineutrino observatory capable of such exposures is under discussion [*Dye et al.*, 2006; *Learned et al.*, 2008].

**14. Conclusions**

This review presents the science and status of geo-neutrino observations, including the prospects for measuring the radioactive power of the planet. Present geo-neutrino detection techniques provide sensitivity to the main heat-producing nuclides $^{238}$U and $^{232}$Th. Techniques presenting directional capability and sensitivity to lower energy geo-neutrinos from $^{235}$U and $^{40}$K require development. Existing observations with limited sensitivity to geo-neutrinos from the mantle constrain radiogenic heating to 15 – 41 TW, assuming a thorium-to-uranium ratio and a homogeneous mantle. This range of acceptable values is comparable to those estimated by geological models (11 – 38 TW) and planetary cooling (13 – 37 TW). Future observations with greater sensitivity to geo-neutrinos from the mantle offer more precise radiogenic heating assessments, approaching 15 % at oceanic locations. More accurate evaluations of the geo-neutrino energy spectrum access the unmeasured thorium-to-uranium ratio, helping to discriminate Earth models if signals have dominant mantle contributions. At continental locations, including the sites for several future observatories, the predicted mantle geo-neutrino contribution is ≈20 % of the total. Observations at oceanic locations far from continents provide measurements of mantle geo-neutrinos that lift the veil of uncertainty obscuring the radioactive power of the planet.

**Acknowledgements**

The author gratefully acknowledges Matt Jackson and three anonymous reviewers for many helpful suggestions and comments. This work was supported in part by National Science Foundation grants through the Cooperative Studies of the Earth's Deep Interior program (EAR 0855838 and EAR 1068097) and by the Hawaii Pacific University Trustees' Scholarly Endeavors Program.

**Table 1.** List of Symbols

| Symbol | Quantity |
|---|---|
| $a$ | isotope or element abundance |
| d | deuteron |
| e | electron |
| $h$ | heat production per unit mass |
| $l$ | antineutrino luminosity |
| $m$ | particle mass |
| n | neutron |
| $n$ | number of particles |
| p | proton |
| $p$ | particle momentum |
| $r$ | position distance |
| $w$ | particle total energy |
| α | alpha particle |
| $\alpha$ | fine structure constant |
| $\varepsilon$ | detector exposure |
| $\lambda$ | nuclear decay constant |
| $\mu$ | molar mass |
| ν | neutrino |
| $\rho$ | mass density |
| $\sigma$ | interaction cross section |
| $\phi$ | geo-neutrino flux |
| $C$ | geo-neutrino event rate to flux conversion factor |
| $E$ | particle total energy |
| $G$ | geological response factor |
| $H$ | heat production |
| $M$ | geological mass |
| $N_A$ | Avogadro's number |
| $P$ | Probability |
| $Q$ | Heat |
| $R$ | geo-neutrino event rate |
| $T$ | particle kinetic energy |
| $V$ | volume |
| Th/U | thorium to uranium abundance ratio |
| CC | cosmochemical model |
| DM | depleted mantle |
| DMM | depleted MORB mantle |
| D" | mantle basement layer |
| GP | geophysical model |
| PM | primitive mantle |

**Table 2.** Parent Nuclide Quantities for Radiogenic Heating and Geo-neutrino Flux

| Isotope | % n. a. | $\mu$(g/mol) | $\lambda(10^{-18}\ s^{-1})$ | $n_{\bar{\nu}_e}$ | $Q$(pJ) | $Q_\nu$(pJ) | $Q_h$(pJ) | $h$(μW/kg) | $l$(kg$^{-1}$μs$^{-1}$) |
|---|---|---|---|---|---|---|---|---|---|
| $^{238}$U | 99.2796 | 238 | 4.916 | 6 | 8.282 | 0.634 | 7.648 | 95.13 | 74.6 |
| $^{235}$U | 0.7204 | 235 | 31.210 | 4 | 7.434 | 0.325 | 7.108 | 568.47 | 319.9 |
| $^{232}$Th | 100.0000 | 232 | 1.563 | 4 | 6.833 | 0.358 | 6.475 | 26.28 | 16.2 |
| $^{40}$K | 0.0117 | 40 | 17.200 | 0.893 | 0.213 | 0.103 | 0.110 | 28.47 | 231.2 |

**Table 3.** Element Specific Heat Generation and Antineutrino Luminosity per Unit Mass

| | $h$(μW/kg) | $l$(kg$^{-1}$μs$^{-1}$) |
|---|---|---|
| Uranium | 98.5 | 76.4 |
| Thorium | 26.3 | 16.2 |
| Potassium | $3.33 \times 10^{-3}$ | $27.1 \times 10^{-3}$ |

**Table 4.** Geological Response Factors and Masses of Earth Reservoirs

| | In. Core | Out. Core | D" | Low. Mantle | Up. Mantle | Crust | Earth |
|---|---|---|---|---|---|---|---|
| $G$ ($10^5$ kg cm$^{-2}$) | 0.19 | 3.86 | 0.29 | 7.32 | 4.16 | 0.20 | 16.03 |
| $M$ ($10^{24}$ kg) | 0.10 | 1.84 | 0.13 | 2.81 | 1.06 | 0.30 | 5.97 |

**Table 5.** Crust Model Information

| ($M_C = 27.9 \times 10^{21}$ kg) | $M$ ($10^{21}$ kg) | $a_U$($10^{-6}$ g/g) | $a_{Th}$($10^{-6}$ g/g) | Th/U | $a_K$($10^{-2}$ g/g) | H(TW) |
|---|---|---|---|---|---|---|
| CC sediments | 0.8 | 2.7 (±21%) | 10.5 (±10%) | 3.9 | 2.4 (±8%) | 0.5 ± 0.1 |
| CC Upper | 6.9 | 2.7 (±21%) | 10.5 (±10%) | 3.9 | 2.4 (±8%) | 4.3 ± 0.6 |
| CC Middle | 7.1 | 1.3 (±31%) | 6.5 (±8%) | 5.0 | 2.0 (±14%) | 2.6 ± 0.4 |
| CC Lower | 6.5 | 0.2 (±80%) | 1.2 (±80%) | 6.0 | 0.5 (±30%) | 0.4 ± 0.3 |
| OC sediments | 0.3 | 2.7 (±21%) | 10.5 (±10%) | 3.9 | 2.4 (±8%) | 0.2 ± 0.0 |
| OC | 6.3 | 0.1 (±30%) | 0.2 (±30%) | 2.2 | 0.1 (±10%) | 0.1 ± 0.0 |
| Bulk Crust | 27.9 | 1.2 (±15%) | 5.0 (±7%) | 4.3 | 1.3 (±7%) | 8.1 ± 0.8 |

**Table 6.** Depleted MORB-Source Mantle Model Information

| | $a_U$(ng/g) | $a_{Th}$(ng/g) | Th/U | $a_K$(μg/g) | $h$(pW/kg) |
|---|---|---|---|---|---|
| Jochum et al. (1983)- best | ~3 | ~6 | 2.0 | ~40 | ~0.6 |
| Jochum et al. (1983)- max | <8 | <16 | 2.0 | <100 | <1.5 |
| Salters and Stracke, (2004) | 4.7 (±30%) | 13.7 (±30%) | 2.9 | 60 (±28%) | 1.0 (±30%) |
| Workman and Hart, (2005) | 3.2 (±18%) | 7.9 (±16%) | 2.5 | 50 (±NA) | 0.7 (±18%) |
| Boyet and Carlson, (2008) | 5.4 (±NA) | 16 (±NA) | 3.0 | 240 (±NA) | 1.2 (±NA) |
| Arevalo and McDonough, (2010) | 8.0 (±20%) | 22 (±20%) | 2.7 | 152 (±20%) | 1.9 (±20%) |

**Table 7.** Primitive Mantle Model Information

| ($M_{SE} = 4.03 \times 10^{24}$ kg) | $a_U$(ng/g) | $a_{Th}$(ng/g) | Th/U | $a_K$(μg/g) | H(TW) |
|---|---|---|---|---|---|
| Chondritic (CI) | 7.8 | 29.8 | 3.8 | 544 | 13 |
| Cosmochemical (CC) | 12 – 22 | 45 – 83 | 3.8 | 140 – 260 | 11 – 21 |
| Geophysical (GP) | 20 – 35 | 77 – 135 | 3.8 | 240 – 420 | 19 – 38 |

**Table 8.** Depleted Mantle Model Information

| ($M_M = 4.00 \times 10^{24}$ kg) | $a_U$(ng/g) | $a_{Th}$(ng/g) | Th/U | $a_K$(μg/g) | R (TNU) | H(TW) |
|---|---|---|---|---|---|---|
| Cosmochemical (CC) | 3.6 – 14 | 10 – 49 | 2.8 – 3.5 | 49 – 170 | 2.7 – 11 | 3.2 – 13 |
| Geophysical (GP) | 12 – 27 | 43 – 100 | 3.6 – 3.7 | 150 – 330 | 9.5 – 26 | 11 – 30 |

**Figure Captions**

Fig. 1. These curves show the antineutrino intensity energy spectra per decay of $^{238}$U, $^{232}$Th, $^{235}$U, and $^{40}$K, which are the main nuclides contributing to terrestrial radiogenic heating and the surface geo-neutrino flux.

Fig. 2. These curves show the total cross sections for the scattering of electron antineutrinos on protons and electrons over the range of energy relevant to geo-neutrinos. For reference the cross section for electron neutrino scattering on electrons is also shown, being about a factor of ~2.4 larger than that for electron antineutrinos.

Fig. 3. These curves show the antineutrino interaction energy spectra per decay of $^{238}$U, $^{232}$Th, $^{235}$U, and $^{40}$K for both proton and electron scattering targets. Note that the electron scattering spectra include all electron recoils, even those down to zero energy.

Fig. 4. These equal area curves show simulated geo-neutrino interaction energy spectra for different values of the thorium-to-uranium abundance ratio Th/U. The spectra assume a detected energy resolution of $\delta E = 7\%\sqrt{E(MeV)}$.

Fig. 5. An upward shift and distortion of the energy spectrum is introduced by using the average oscillation probability. The shift and distortion are more pronounced at a continental site than at an oceanic site. Simulations consider a continental planet with 45-km thick crust and an oceanic planet with 8-km thick crust. The continental observatory is 2-km below the surface, while the oceanic observatory is 0.1-km above the crust.

fig. 6. This figure compares estimates of heat-producing, refractory element (U, Th) abundances in chondrites. The analyses in this review use the estimate of PO 2003 [*Palme and O'Neill*, 2003], which is consistent with RJ 1993 [*Rocholl and Jochum*, 1993] and MS 1995 [*McDonough and Sun*, 1995]. The black dotted line follows constant Th/U ratio equal to 3.8.

Fig. 7. Panel (a) shows the depleted mantle mass fraction as a function of primitive mantle enrichment of refractory heat-producing elements (U, Th), assuming various estimates (WH [*Workman and Hart*, 2005]; SS [*Salters and Stracke*, 2004]; AM [*Arevalo and McDonough*, 2010]) of heat-producing element abundances in depleted mid-ocean ridge basalt (MORB) source mantle (DMM) for the depleted mantle. Panel (b) shows the variation of heat-producing element (U, Th, K) enrichment in an enriched layer at the base of the mantle as a function of primitive mantle enrichment of refractory heat-producing elements (U, Th), assuming various estimates (WH [*Workman and Hart*, 2005]; SS [*Salters and Stracke*, 2004]; AM [*Arevalo and McDonough*, 2010]) of heat-producing element abundances in depleted mid-ocean ridge basalt (MORB) source mantle (DMM) for the depleted mantle.

Fig. 8. Panel (a) shows the variation of mantle geo-neutrino rate as a function of primitive mantle enrichment of refractory heat-producing elements (U, Th) for a homogeneous distribution (maximum rate) and for DMM (SS) overlying an enriched layer (D") at the base of the mantle (minimum rate). The green shaded area between the curves specifies the allowed parameter space. For comparison, the combined geo-neutrino result (KL+BX) with uncertainty is plotted over the allowed space. Panel (b) shows the variation of mantle geo-neutrino Th/U ratio as a function of primitive mantle enrichment of refractory heat-producing elements (U, Th) for a homogeneous distribution (maximum rate) and for DMM (SS) overlying an enriched layer at the base of the mantle (minimum rate). The green shaded area between the curves specifies the allowed parameter space.

Fig. 9. This figure compares the radiogenic heating implied by the geo-neutrino observations of KamLAND (KL) and Borexino (BX) with the ranges of radiogenic heating predicted by the cosmochemical (CC) and geophysical (GP) models and with the surface heat flow. Minimum implied heating results from a homogeneous distribution of heat-producing elements in the mantle. Maximum implied heating results from an enriched layer of heat-producing elements at the core-mantle boundary overlain by a homogeneous mantle with depleted MORB-source mantle composition. The combined results employ a weighted average, assuming both observations sample identical mantle signals.

Fig. 10. This figure plots the predicted antineutrino energy spectra for Sudbury, Canada, showing the contributions from geo-neutrinos (U and Th) and nuclear reactors. The spectra assume a detected energy resolution of $\delta E = 7\% \sqrt{E(MeV)}$ .

Fig. 11. This figure plots expected geo-neutrino signals (solid lines) with systematic errors (dotted lines) expected at an oceanic (blue), existing (black), and continental site (brown) as a function of radiogenic heating. Assuming a terrestrial radiogenic heating of 20 TW, the systematic uncertainty introduces error (dashed lines) of 15 % at the oceanic site, 30 % at an existing site, and 40 % at a continental site.

Figure 1

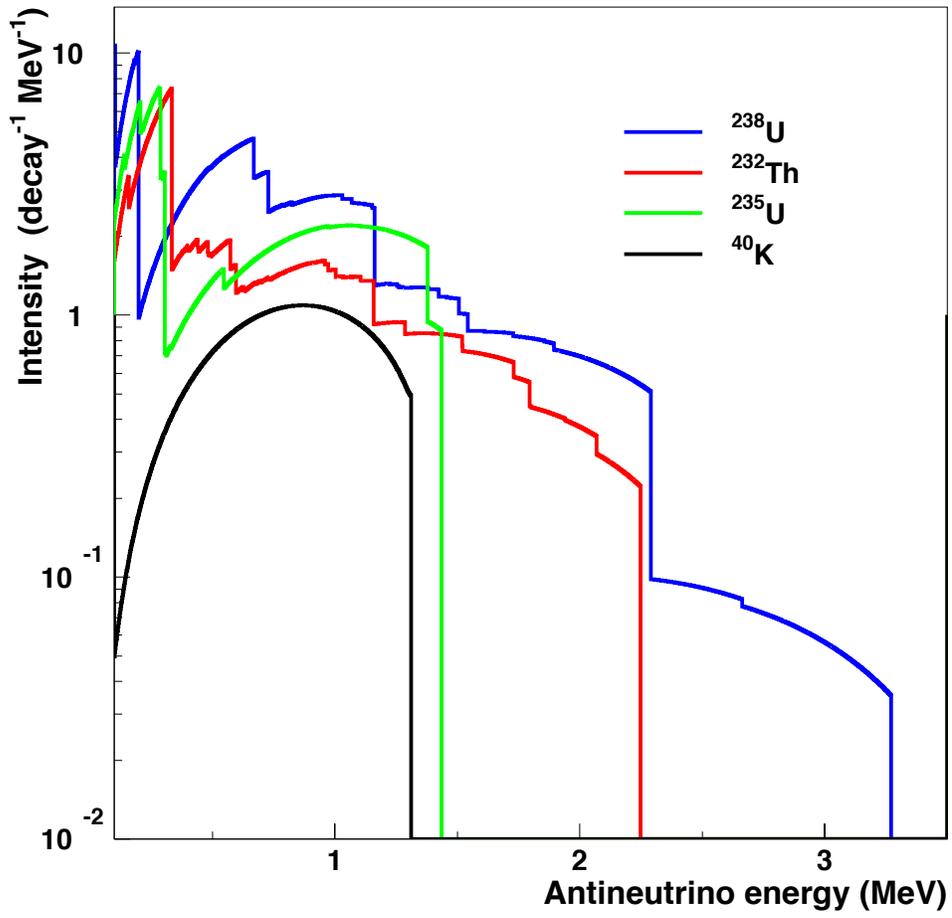

Figure 2

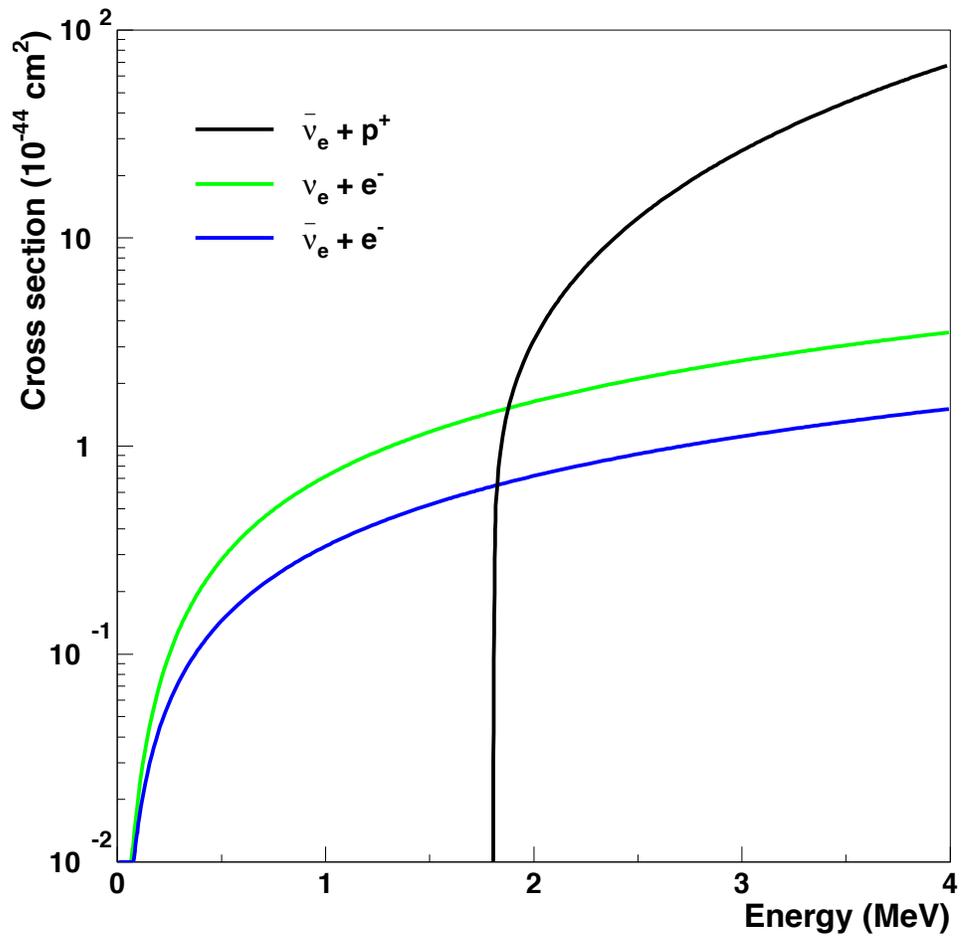

Figure 3

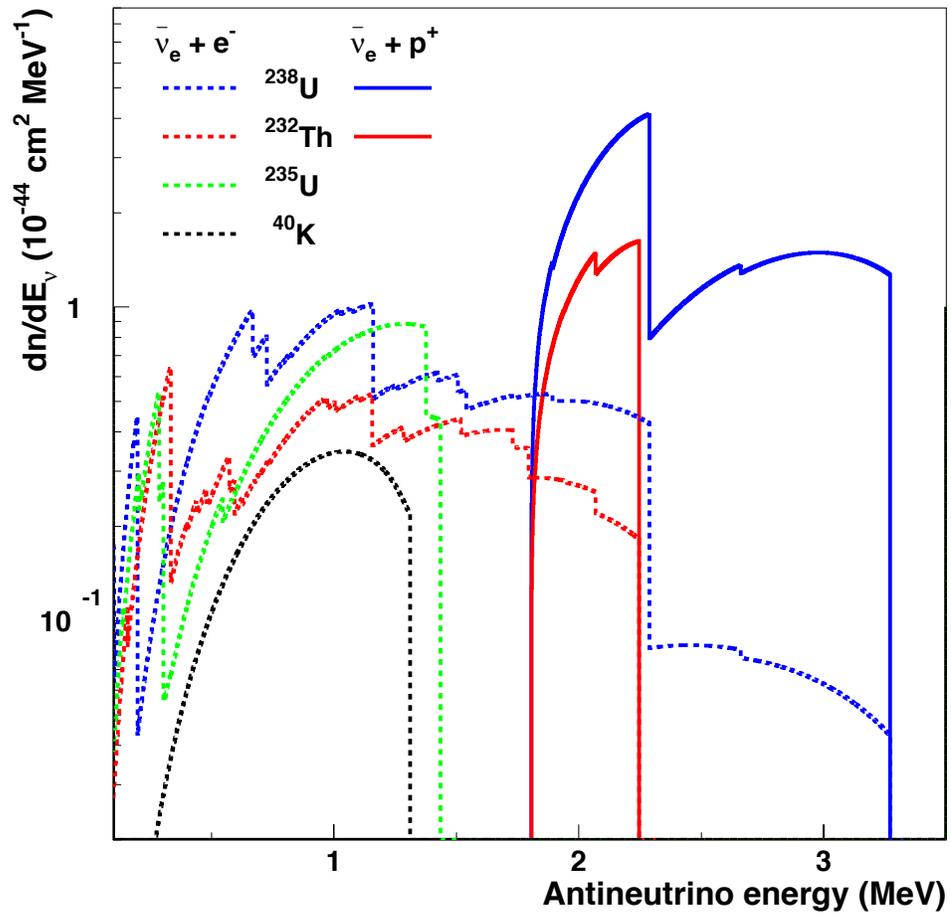

Figure 4

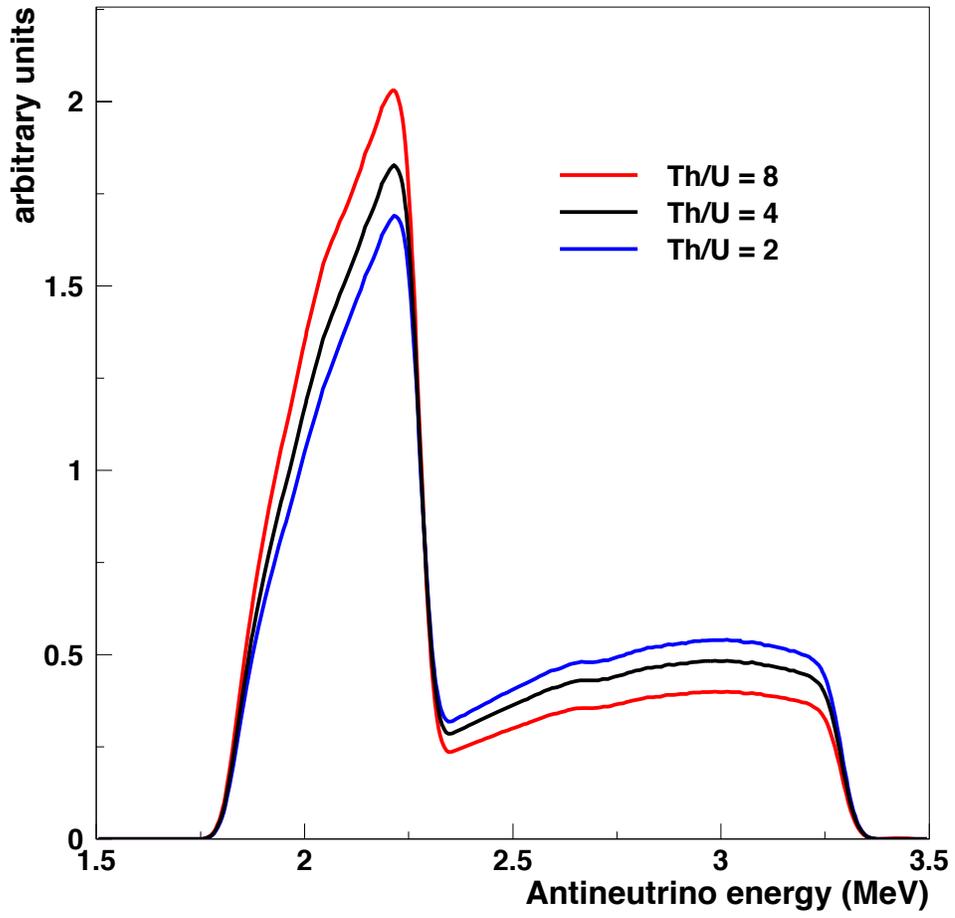

Figure 5

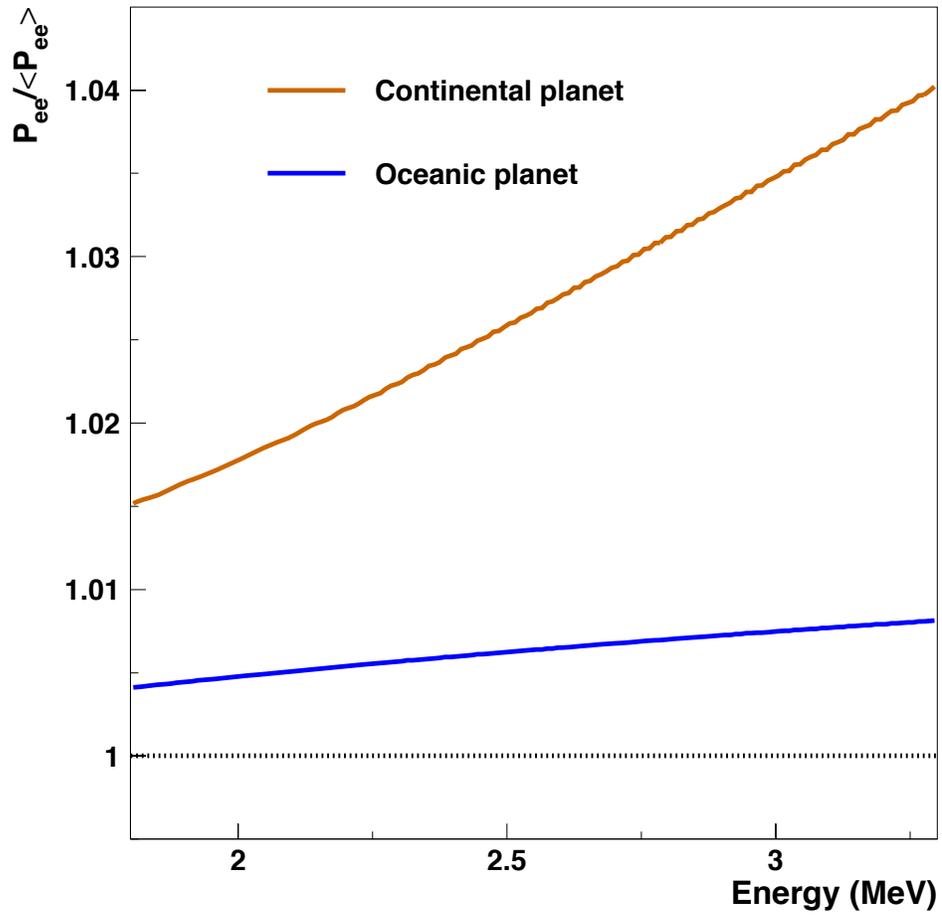

Figure 6

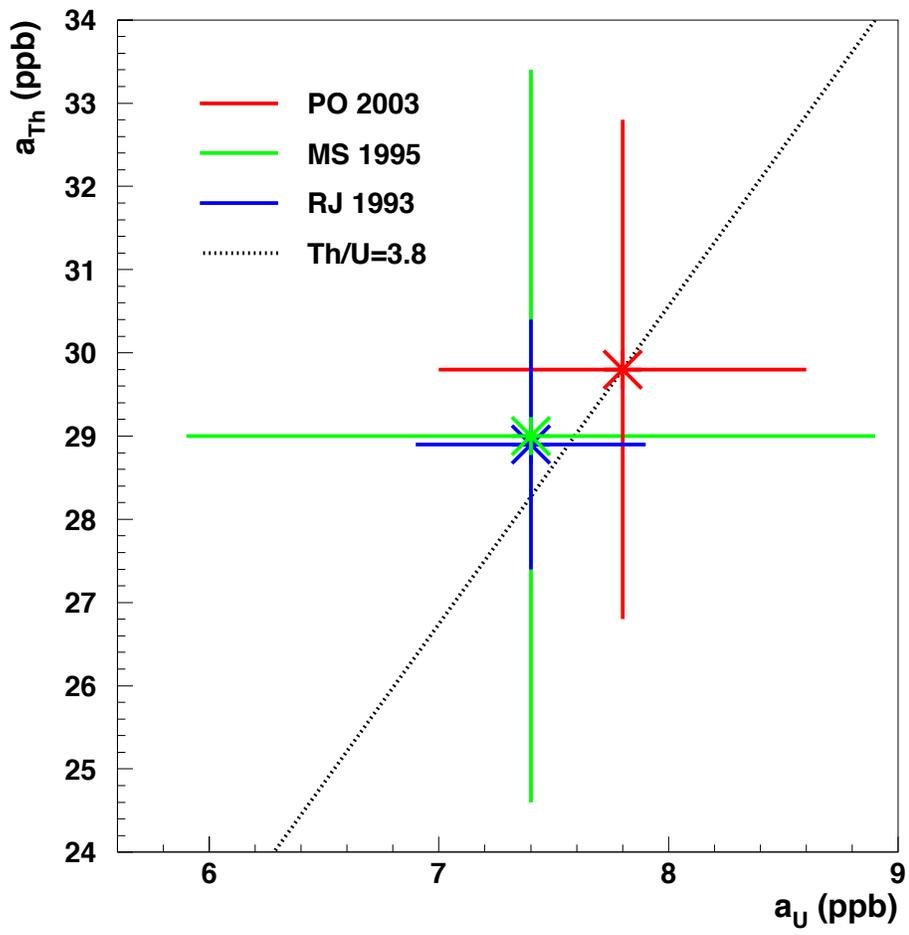

Figure 7a

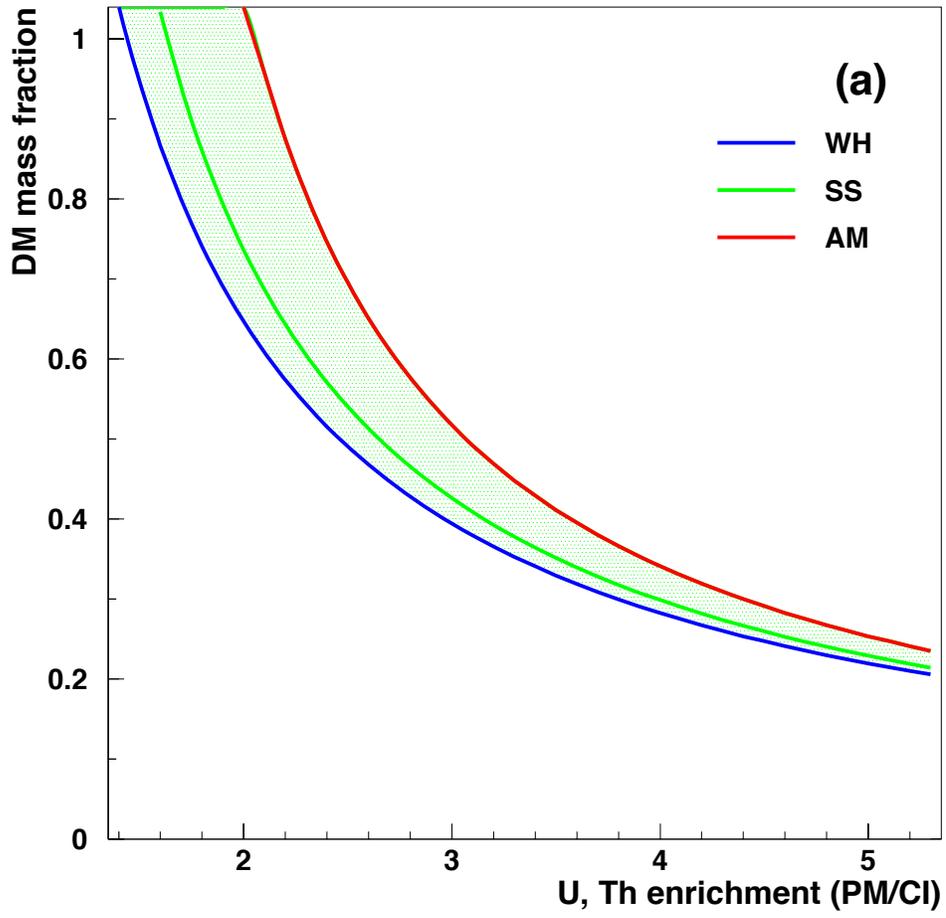

Figure 7b

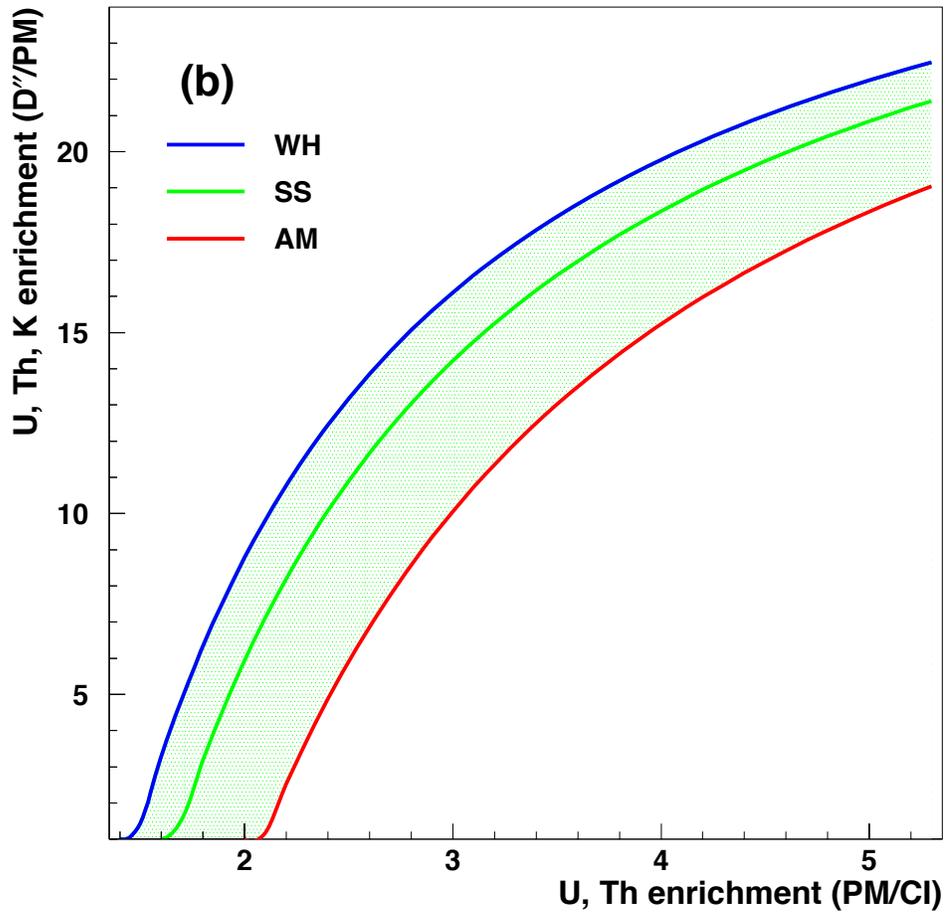

Figure 8a

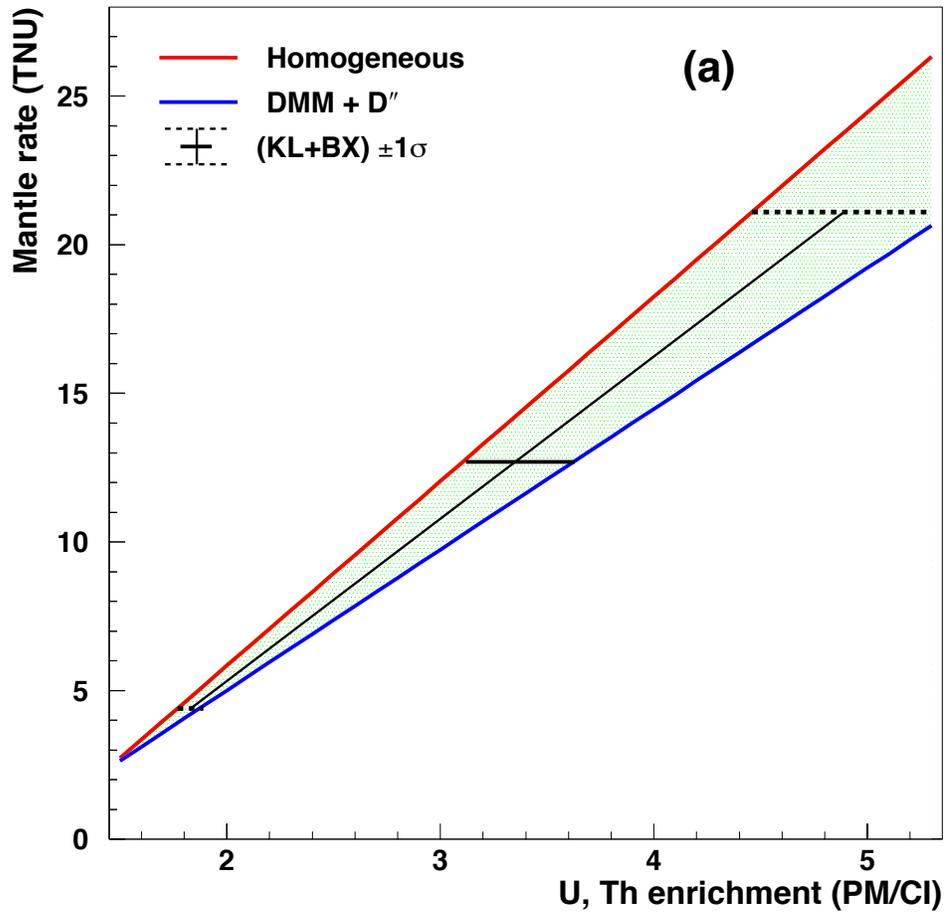

Figure 8b

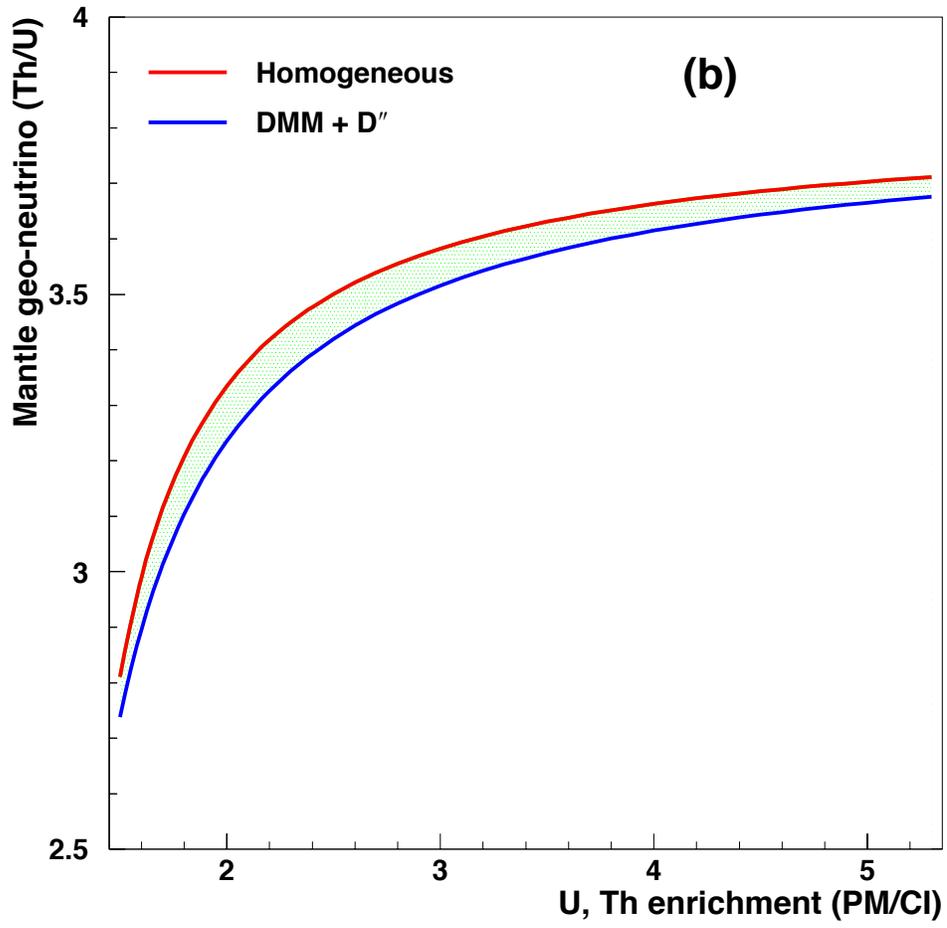

Figure 9

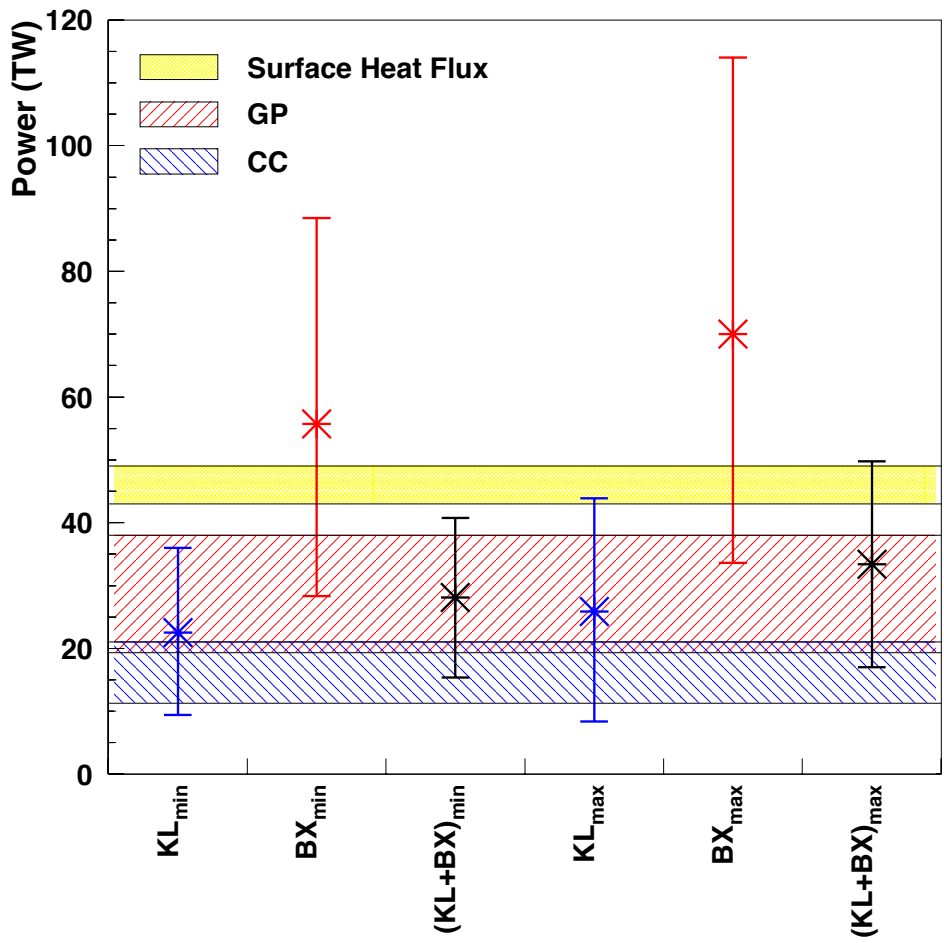

Figure 10

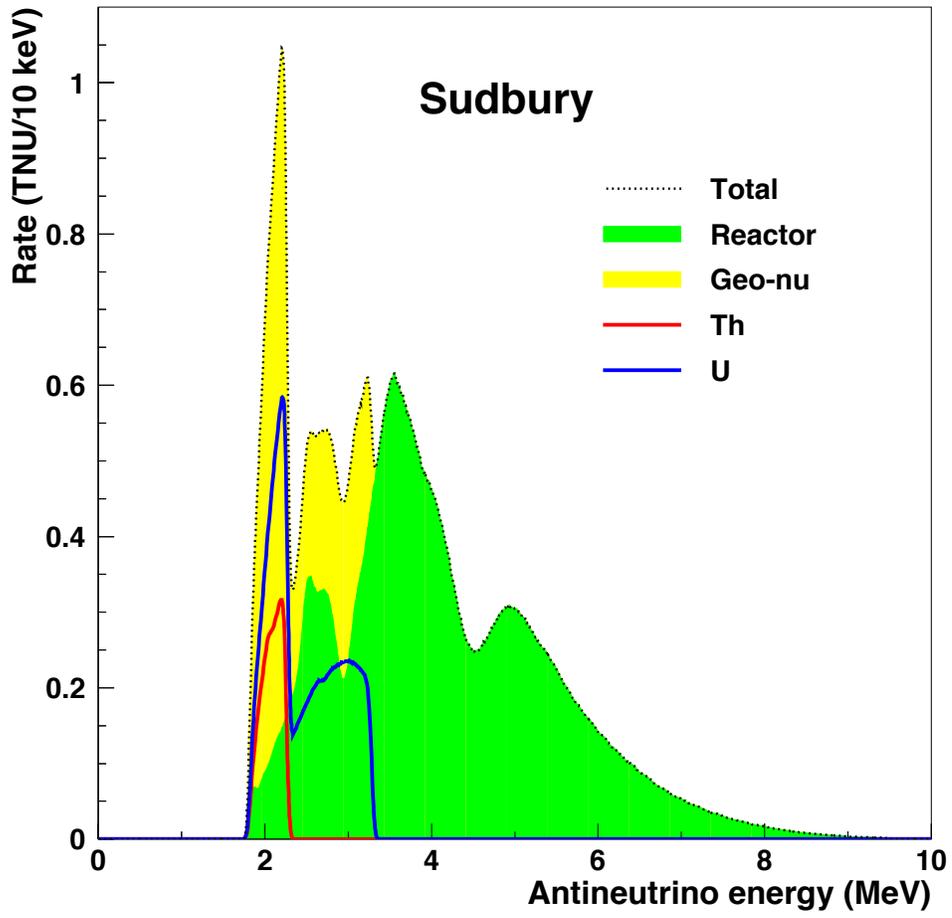

Figure 11

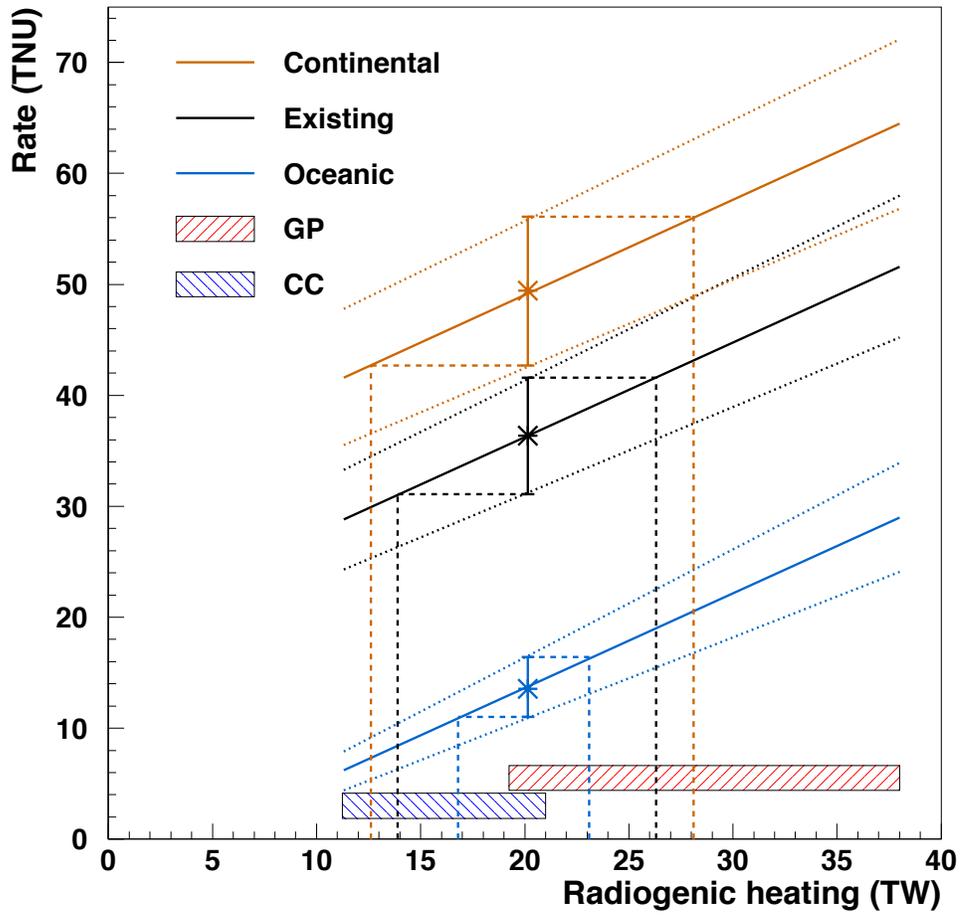

48